\documentclass[aip,reprint]{revtex4-1}

\usepackage[utf8]{inputenc} 
\usepackage[T1]{fontenc}    
\usepackage{hyperref}       
\usepackage{url}            
\usepackage{booktabs}       
\usepackage{amsfonts}       
\usepackage{nicefrac}       
\usepackage{microtype}      
\usepackage{graphicx}
\usepackage{lipsum}
\usepackage{comment}
\usepackage[caption=false]{subfig}
\usepackage{textcomp}
\usepackage{soul}           
\usepackage{amsmath}
\usepackage{listings}
\usepackage{chngcntr}
\usepackage{enumitem}
\usepackage[table,xcdraw]{xcolor}
\usepackage{booktabs}
\usepackage{multirow}
\usepackage{hhline}
\usepackage{amssymb}
\usepackage{siunitx}
\usepackage{bibunits}
\defaultbibliography{references}      
\defaultbibliographystyle{apsrev4-1}      

\usepackage{euler}

\definecolor{codegreen}{rgb}{0,0.6,0}
\definecolor{codegray}{rgb}{0.5,0.5,0.5}
\definecolor{codepurple}{rgb}{0.58,0,0.82}
\definecolor{backcolour}{rgb}{0.95,0.95,0.92}

\lstdefinestyle{mystyle}{
    backgroundcolor=\color{backcolour},   
    commentstyle=\color{codegreen},
    keywordstyle=\color{magenta},
    numberstyle=\tiny\color{codegray},
    stringstyle=\color{codepurple},
    basicstyle=\ttfamily\footnotesize,
    breakatwhitespace=false,         
    breaklines=true,                 
    captionpos=b,                    
    keepspaces=true,                 
    numbers=left,                    
    numbersep=5pt,                  
    showspaces=false,                
    showstringspaces=false,
    showtabs=false,                  
    tabsize=2
}
\let\oldtextbf=\textbf
\renewcommand*{\textbf}[1]{\ifmmode\mathbf{#1}\else\oldtextbf{#1}\fi}
\renewcommand*{\phi}[0]{\varphi}

\newcommand{\paragraphtitle}[1]{\textsf{\textbf{\small {#1}}}}

\draft 

\begin{document}

\title{Ceci n’est pas un committor, yet it samples like one: efficient sampling via approximated committor functions.}



\author{Enrico Trizio}
\email[]{enrico.trizio@iit.it}
\affiliation{$^1$Atomistic Simulations, Italian Institute of Technology, 16156 Genova, Italy}

\author{Giorgia Rossi}
\affiliation{$^1$Atomistic Simulations, Italian Institute of Technology, 16156 Genova, Italy}

\author{Michele Parrinello$^*$}
\email[]{michele.parrinello@iit.it}
\affiliation{$^1$Atomistic Simulations, Italian Institute of Technology, 16156 Genova, Italy}


\date{\today}

\begin{abstract}
Atomistic simulations are widely used to investigate reactive processes but are often limited by the rare event problem due to kinetic bottlenecks.
We recently introduced an enhanced sampling approach based on the committor function, machine-learned following a variational principle. 
This method combines a transition-state-oriented bias potential, expressed as a functional of the committor, with a metadynamics-like bias along a committor-based collective variable, enabling uniform exploration of reaction pathways. 
In its original formulation, the committor is represented by a neural network that takes physical descriptors as input and is trained by minimizing a functional involving gradients with respect to atomic coordinates, which can be computationally demanding in some cases.
Here, we propose a simplified learning criterion formulated entirely in the descriptor space, which bypasses the need for explicit and costly coordinate gradients and provides a relaxed upper bound to the original variational principle.
Although this approach does not formally target the \textit{exact} committor, we show that it retains robust sampling performance while significantly reducing computational costs, thus enabling the study of processes that would be practically unfeasible using the original formulation.
\end{abstract}


\maketitle 
\begin{bibunit}

\section{Introduction}
    Atomistic simulations are one of the pillars of modern science, enabling the study of the microscopic mechanisms underlying reactive processes in fields as diverse as biophysics, materials science, and chemistry.
    However, the scope of such methods is still limited by the timescales accessible to standard simulation algorithms, which are typically much shorter than those of most reactive processes, making them difficult to observe in simulations.~\cite{frenkel2001understanding}
    This is due to the so-called rare event problem, which arises when metastable states are separated by kinetic bottlenecks. 
    To address this issue, over the last few decades, several enhanced sampling methods aimed at facilitating the study of rare events have been developed.~\cite{Henin2022enhanced, zhu2025enhanced}
    
    Recently, we introduced a promising method~\cite{kang2024computing, trizio2025everything} based on the calculation of the committor function combined with the On-the-fly  Probability Enhanced Sampling (OPES) method.~\cite{invernizzi2020rethinking, trizio2024advanced}
    This approach has demonstrated unique capabilities in uniformly sampling entire reaction pathways and has several merits: it is based on a physics-grounded variational principle, it enables exhaustive sampling of the transition state (TS) region, and it provides a powerful tool to reveal the reaction process in remarkable detail.
    In the method section, we summarize the key concepts of this approach, while a more extensive discussion can be found in the original papers.~\cite{kang2024computing, trizio2025everything}
    
    The central quantity of this approach is the committor function $q(\textbf{x})$, which, for the transition between two states $A$ and $B$, gives the probability that a configuration  $\textbf{x}$ evolves to state $B$ before having first passed by $A$.~\cite{weinan2010transition}
    Owing to this probabilistic description, the committor is widely considered to provide the ideal description of a reactive process, and due to its many attractive features, much work has been done to determine it.~\cite{bolhuis2000reaction,ma2005automatic,berezhkovskii2005one,peters2006obtaining,e2006towards,li2019computing,khoo2019solving,rotskoff2022active,he2022committor,chen2023committor,chen2023discovering,jung2023machine,mitchell2024committor}
    However, computing $q(\textbf{x})$ is challenging, as it requires extensive sampling of the TS region.
    Traditionally, this was obtained either from long simulations~\cite{krivov2021nonparametric} or via transition path sampling.~\cite{bolhuis2002throwing,bolhuis2018nested,Dellago2006,jung2023machine}
    Recently, we have shown that once the $A$ and $B$ states are known, the committor can be computed via a self-consistent iterative procedure.~\cite{kang2024computing,trizio2025everything}
    Our approach is based on a variational principle first pioneered by Onsager\cite{onsager1938recombination} and later framed into the formalism of Kolmogorov's backward equation.~\cite{kolmogoroff1931analytischen}
    This principle states that, under the assumption of overdamped Langevin dynamics, the committor can be determined by imposing the boundary conditions $q(\textbf{x}_{i\in A})=0$ and $q(\textbf{x}_{i\in B})=1$ and minimizing the functional
    \begin{equation*}
        K[q(\textbf{x})]=\langle \vert \nabla_{\textbf{u}}  q(\textbf{x})\vert^2 \rangle
    \end{equation*}
    in which the gradients $\nabla_{\textbf{u}}$ are computed with respect to the mass-scaled atomic coordinates and $\langle \cdot \rangle$ denotes the average over the equilibrium Boltzmann ensemble. 
    
    In Ref.~\citenum{trizio2025everything}, the committor was expressed as a feed-forward neural network $q_{\boldsymbol{\theta}} (\textbf{d}(\textbf{x}))$, where $\boldsymbol{\theta}$ are learnable parameters. 
    The input of the neural network is a set of $N_d$ physical descriptors $\textbf{d}(\textbf{x})$ which provide a convenient way to enforce the system's physical symmetries and favor interpretability. 
    The model parameters are optimized using as a loss function the $K[q(\textbf{x})]$ functional supplemented by the above-mentioned boundary conditions. 
  
    Due to its  ability to extensively and uniformly sample the reactive process, our approach  has proven robust and successful in applications that range from biophysics~\cite{das2025machine, berselli2026, trizio2025everything} to chemistry and materials science.~\cite{trizio2025everything, kang2024computing, deng2026fluctuations}
    However, in some cases, training the committor can be rather expensive.
    The main numerical source of inefficiency is often the calculation of the gradients relative to the atomic coordinates in the variational functional. 
    The cost of computing these derivatives can be rather high when using complex descriptors that depend on a large number of atoms.~\cite{deng2026fluctuations, dietrich2024graph}
    
    The aim of this paper is to reduce the computational cost of this part of the calculation while retaining its overall sampling efficiency.  
    To this effect, we replace the original functional with a new one that depends only on the derivatives of the committor relative to the descriptors
      \begin{equation*}
         \tilde{K}[ q(\textbf {d}(\textbf{x}))]=\langle \vert \nabla_{\mathbf{d} } q(\textbf {d}(\textbf{x}))\vert^4\rangle_{U(\textbf{x})}  
     \end{equation*}
    This choice can be motivated with the help of the Cauchy-Schwarz inequality, showing that 
    $\tilde{K}[ q(\textbf {d}(\textbf{x}))]$ is an upper bound to ${K}[ q(\textbf {d}(\textbf{x}))]^2$.

    Even if this approach does not lead to computing the \textit{true} committor, we show that it provides a reliable and economical sampling tool across a number of paradigmatic examples, including the conformational equilibrium of alanine dipeptide, an intramolecular proton transfer reaction, a binding problem, and the crystallization of silicon from melt, some of which would have been unfeasible with the original approach.
    We hope that this clarifies the title of the manuscript, which is inspired by the painting \textit{Ceci n'est pas une pipe} (This is not a pipe) by the surrealist painter René Magritte, which, ironically, represents a pipe.
    
\section{Methods}

In the following, we start by giving an overview of the whole iterative procedure in Sec.~\ref{sec:summary}, highlighting the key differences with Ref.~\citenum{trizio2025everything}.
Then, we discuss in detail all the components of such a procedure. 
Specifically, in Sec.~\ref{sec:background}, we recall the general features of the committor function and the main concepts of the original approach of Ref.~\citenum{trizio2025everything}.
Then, in Sec.~\ref{sec:reformulation}, we present in detail the modifications to the variational loss function used for the optimization of the approximated committor models.  

\subsection{Summary of the self-consistent iterative procedure.} 
\label{sec:summary}
Here, we outline the steps of the self-consistent iterative procedure proposed here for the learning of approximated committor functions.
Such a procedure closely resembles the original recipe of Ref.\citenum{trizio2025everything}, except for the use of a different loss function that leads to an approximated committor model, as detailed in the following sections.
\begin{itemize}
    \item\textbf{Step 1:} At iteration $n$, the committor model $q_{\mathbf{\theta}}^{n}(\textbf{x})$ is trained on the dataset of configurations $\textbf{x}^n$ and weights $\textbf{w}^n$ accumulated up to iteration $n$.
    In the first iteration, i.e., $n=0$, the dataset is either composed of data from previous calculations or solely of labeled configurations from the metastable basins.
    The optimization of the model is performed by minimizing the modified loss functions in Eqs.~\ref{eq:lbound} and \ref{eq:lvar}.
    \item \textbf{Step 2:} The committor estimate is used to perform enhanced sampling simulations under the combined action of the Kolmogorov bias $V_K^n$ and the OPES bias $V_{\text{OPES}}^n$.
    The latter is applied along the committor-related CV $z^n(\textbf{x})$.
    \item\textbf{Step 3:} The sampled configurations are used to update the training dataset after being reweighted according to the effective bias $V_{\text{eff}}^n = V_{K}^n + V_{\text{OPES}}^n$ with weights given by $w_i^{n} =\frac{e^{\beta\,V_{\mathrm{eff}}^{n}(\textbf{x}_{i})}} {\left\langle e^{\beta\,V_{\mathrm{eff}}^{n}(\textbf{x})}\right\rangle_{U_{\mathrm{eff}}^{n}}}$.
\end{itemize}
The above steps are iterated to refine the committor estimate and improve sampling until convergence is reached.

\subsection{Background}
\label{sec:background}

\paragraphtitle{Committor function.} 
As mentioned in the introduction, given two metastable states $A$ and $B$, the committor $q(\textbf{x})$ is a function of the atomic coordinates $\textbf{x}$ which returns the probability that a trajectory initiated from configuration $\textbf{x}$ will reach state $B$ before passing through $A$. 
As a consequence, the committor resembles a step-like function, being nearly constant in the metastable basins, i.e., $q(\textbf{x}_{i\in A}) \simeq 0$ and $q(\textbf{x}_{i\in B}) \simeq 1$, and displaying significant gradients only over the TS region. 
Due to its probabilistic nature, the committor is arguably the most principled description of a reactive process, and it is believed to be an ideal reaction coordinate.~\cite{weinan2010transition}

However, determining the committor is challenging. 
One possibility is to leverage a variational approach pioneered by Onsager and later formalized using the Kolmogorov backward equation, which states that, given the boundary conditions $q(\mathbf{x}) = 0$ for $\mathbf{x} \in A$ and $q(\mathbf{x}) = 1$ for $\mathbf{x} \in B$, the committor can be obtained by minimizing the functional
    \begin{equation}
        K[q(\textbf{x})] = \langle |\nabla_{\textbf{u}} q(\textbf{x})|^2 \rangle
        \label{eq:kfunctional}
    \end{equation}
where $\langle \cdot \rangle$ denotes the average on the Boltzmann ensemble associated with the interatomic potential $U(\textbf{x})$, and $\nabla_{\textbf{u}}$ the gradient with respect to the mass-weighted coordinates $\textbf{u}$.
Unfortunately, even if Eq.~\ref{eq:kfunctional} could be evaluated through sampling, its practical application remains challenging due to the strong localization of $|\nabla q(\textbf{x})|$ on the TS region, which is highly unfavored in simulations.

\paragraphtitle{Kolmogorov bias.} 
To overcome the sampling limitations associated with the TS in rare events, in Ref.~\citenum{kang2024computing}, we introduced the Kolmogorov bias potential $V_K(\textbf{x})$ to stabilize the TS region and favor its sampling.
This TS-oriented bias is defined as a functional of the committor, exploiting the strong localization of its gradients on the TS
    \begin{equation}
        V_K (\textbf{x})= -\frac{\lambda}{\beta} \log (|\nabla q(\textbf{x})|^2 + \epsilon)
            \label{eq:kbias}
    \end{equation}
where $\beta=\frac{1}{k_B T}$ is the inverse temperature and $\lambda$ a parameter regulating the magnitude of the bias.
In turn, $\epsilon$ is a regularization term that can either simply avoid numerical instabilities (i.e., $\epsilon \ll 1$) or filter out contributions with negligible $|\nabla q(\textbf{x})|$ (i.e., $\epsilon=1$).
In this work, as a safety measure, we opted for the latter option to avoid possible artifacts in the metastable regions.

In practice, when applying $V_K(\textbf{x})$ in enhanced sampling calculations, one can equivalently compute it considering the gradients with respect to the atomic coordinates or to the input descriptors, as discussed in Ref.~\citenum{kang2024computing}.
As the second possibility is simpler to implement and computationally cheaper, it is to be preferred. 

The bias $V_K(\textbf{x})$ is attractive where $|\nabla q(\textbf{x})|$ is large, thus, in the biased energy landscape $U_K(\textbf{x}) = U(\textbf{x}) + V_K(\textbf{x})$, the TS region is stabilized and its sampling is favored. 
Furthermore, $U_K(\textbf{x})$ can be used to define the Kolmogorov probability distribution as 
    \begin{equation}
        p_{K}(\textbf{x})= \frac{e^{-\beta U_{K}(\textbf{x})}}{Z_{K}} \quad\text{with}\quad Z_{K} =\int e^{-\beta\,U_{K}(\textbf{x})}\,\mathrm{d}\textbf{x}
        \label{eq:kprob}
    \end{equation}
which offers a practical way to identify the transition state ensemble (TSE).~\cite{kang2024computing,trizio2025everything,das2025machine,berselli2026}

\paragraphtitle{Machine learning the committor.} 
In Ref.~\citenum{trizio2025everything}, we leveraged the variational principle of Eq.~\ref{eq:kfunctional} to optimize a machine-learning-based parametrization of the committor through a self-consistent iterative procedure.
During such a procedure, cycles of (enhanced) sampling and training are alternated to progressively improve both the committor estimate and the sampling until convergence is reached.

As far as the learning part is concerned, the committor is represented as a feedforward neural network (NN) $q_{\boldsymbol{\theta}}(\textbf{x}) = f_{\boldsymbol{\theta}}(\textbf{d}(\textbf{x}))$ with trainable parameters $\boldsymbol{\theta}$, taking as input a set of physical descriptors $\textbf{d}(\textbf{x})$.
These descriptors, in turn, are designed to capture the essential degrees of freedom of the system, enforce required symmetries, and incorporate prior physical information.
In addition, in order to impose the correct functional form on the output, we apply a sigmoid-like activation function $\sigma$ to the raw output $z(\textbf{x})$ of the NN, thus obtaining the committor as
    \begin{equation}
        q(\mathbf{x}) = \sigma\bigl(z(\mathbf{x})\bigr)
        \label{eq:sigma}
    \end{equation} 
The network parameters are optimized by minimizing a loss function composed of two contributions.
One is the \textit{boundary loss} term, which imposes the correct boundary conditions, and is computed on a labeled dataset of configurations from the initial and final states.
The other is the \textit{variational loss} term, which is based on the variational functional of Eq.~\ref{eq:kfunctional} and is evaluated on the data obtained via enhanced sampling simulations.
 
The simulations run during the iterative procedure are driven by two complementary bias potentials.
On the one hand, the TS-oriented Kolmogorov bias $V_K(\textbf{x})$ (see Eq.~\ref{eq:kbias}) is employed to stabilize the TS region to favor its sampling.
On the other hand, a metadynamics-like bias, based on the On-the-fly probability enhanced sampling~\cite{invernizzi2020rethinking, trizio2024advanced} (OPES) framework, is used to smooth out the free energy landscape, thereby favoring transitions between states and ergodic sampling.
In practice, the OPES bias uses $z(\textbf{x})$ as a CV rather than $q(\textbf{x})$, since it encodes the same information as $q(\textbf{x})$, but has a smoother behavior that makes it more suitable for this application.

\subsection{Reformulation of variational loss}
\label{sec:reformulation}

While the procedure summarized above provides a reliable framework that has already been successfully applied to a range of systems,~\cite{kang2024computing,trizio2025everything,das2025machine,deng2026fluctuations,berselli2026,kang2026without} it still presents some practical drawbacks in certain cases.
Indeed, during training, the main computational cost arises from the minimization of the variational term, which requires computing the gradients $\nabla_\textbf{u}q_{\boldsymbol{\theta}}(\textbf{x})$ with respect to the mass-scaled atomic positions $\textbf{u}$, which in many applications may become particularly demanding.
Specifically, using automatic differentiation packages,~\cite{paszke2019pytorch} the computational load of the training is proportional to i) the number of descriptors used, ii) their complexity, and iii) the number of atoms involved in the calculation of such descriptors.
Even if this is not a conceptual limitation and strategies to reduce the overload can be devised, in practice, this can introduce an extra layer of complexity.
This motivated us to seek a reformulation of our learning approach that would reduce the computational burden and simplify the set up protocol.
    
To this aim, we start by splitting $\nabla_\textbf{u} q_{\boldsymbol{\theta}}(\textbf{x})$ into two contributions using the chain rule of differentiation
    \begin{equation}
        \nabla_\textbf{u} q_{\boldsymbol{\theta}}(\textbf{x}) = 
        \nabla_\textbf{u} d(\textbf{x}) \,
        \nabla_\textbf{d} q_{\boldsymbol{\theta}}(\textbf{x})
    \end{equation}
thus decoupling the derivatives with respect to mass-weighted positions $\nabla_\textbf{u}$ and descriptors $\nabla_\textbf{d}$.
Notably, only the latter contribution depends on the NN model's variational parameters $\boldsymbol{\theta}$, and thus changes during training, whereas the former does not change as the model is updated, as it is fully determined by the choice of descriptors. 
In addition, computation of the $\nabla_\textbf{u} d(\textbf{x})$ matrix can be both computationally expensive, as its size scales with the number of atoms involved in the descriptors calculation, and inefficient, as only a few atoms contribute to each descriptor, thus typically making it extremely sparse.

Both these considerations suggest that a learning criterion can be devised based only on $\nabla_\textbf{d} q_{\boldsymbol{\theta}}(\textbf{x})$, trading a reasonable approximation for potentially large computational savings. 
Besides this somewhat naive intuition, such an alternative approach can also be formally backed by finding an upper bound to the original variational principle, as we now proceed to discuss.
For simplicity of notation, from now on, we omit the $\boldsymbol{\theta}$ subscript when referring to the parametrization of the committor, i.e., $q_{\boldsymbol{\theta}}(\textbf{x}) \rightarrow q(\textbf{x})$.

Making the dependence on the input descriptors explicit, the Kolmogorov functional of Eq.~\ref{eq:kfunctional} can be expressed as 
    \begin{equation}
        K[q(\textbf{d}(\textbf{x}))] = \langle |\nabla_{\textbf{u}} q(\textbf{d}(\textbf{x}))|^2\rangle
    \end{equation}
which, after simple manipulations, fully detailed in the supplementary material, can be written as 

\begin{equation}
      K[q(\textbf{d}(\textbf{x}))] =  \langle g_{i,j}k_{i,j}\rangle
      \label{eq:functional_split}
    \end{equation}
where 
   \begin{equation}
        g_{i,j}= \frac{\partial d_i }{\partial u_{l,n}} \frac{\partial d_j }{\partial u_{l,n}} \qquad
        k_{i,j}= \frac{\partial q }{\partial d_{i}} \frac{\partial q}{\partial d_{j}}
        \label{eq:gk_terms}
    \end{equation} 
In the above equation, the Einstein summation convention over repeated indices is implied, and the indices $l$ and $n$ refer, respectively, to the $N$ atom coordinates and their Cartesian components, while $i$ and $j$ run over the $N_d$ descriptors.

This reformulation allows the Cauchy-Schwarz inequality to be applied, leading to
    \begin{equation}
        \langle g_{i,j}k_{i,j}\rangle^2 \leq  \langle g_{i,j}g_{j,i}\rangle  \langle k_{i,j}k_{j,i}\rangle
        \label{eq:cauchy_inequality}
    \end{equation}
In this expression, the purely geometric term $\langle g_{i,j}g_{j,i}\rangle$ is decoupled from the model-dependent term $\langle k_{i,j}k_{j,i}\rangle $ that only depends on the committor derivatives with respect to the descriptors. 
Thus, the latter can be used as a new and simplified objective function for model optimization. 
Remarkably, this formulation avoids explicitly evaluating the descriptor gradients with respect to the atomic coordinates, leading to a significantly reduced computational cost.

In addition, using elementary algebra, computing this term can be further simplified.
Specifically, defining the vector
    \begin{equation}
        \mathbf{h} =
        \begin{pmatrix}
        \frac{\partial q}{\partial d_1} \\
        \vdots \\
        \frac{\partial q}{\partial d_{N_d}}
        \end{pmatrix}
        \label{eq:vector_h}
    \end{equation}
and writing matrix $\textbf{K}$ as the outer product $\textbf{K} = \mathbf{h}\,\mathbf{h}^\mathrm{T}$, we can obtain a simpler expression for the new variational component of the loss function as
    \begin{equation}
        \langle k_{ij} k_{ji} \rangle =
        \langle \operatorname{tr}(\textbf{K}^2) \rangle = 
        \langle \| \textbf{h} \| ^4 \rangle
        \label{eq:trace_squared_K}
    \end{equation}
which depends only on the derivatives of the committor with respect to the descriptors. 

During training, Eq.\ref{eq:trace_squared_K} is used as a variational loss $L_v$, which is evaluated on a dataset of $N_v$ configurations from enhanced sampling simulations according to
    \begin{equation}
        L_v = \frac{1}{N_v} \sum_i^{N_v} w_i \left| \nabla_{\mathbf{d}} q(\mathbf{x}_i) \right|^4
            \label{eq:lvar}
    \end{equation}
where $w_i =\frac{e^{\beta\,V_{\mathrm{eff}}(\textbf{x}_{i})}} {\left\langle e^{\beta\,V_{\mathrm{eff}}(\textbf{x})}\right\rangle_{U_{\mathrm{eff}}}}$ are statistical weights to reweigh each configuration $\mathbf{x}_i$ by the total bias $V_{\text{eff}} = V_{K} + V_{\text{OPES}}$ to its corresponding Boltzmann probability.

    \begin{figure}[b!]
        \centering
        \includegraphics[width=1\linewidth]{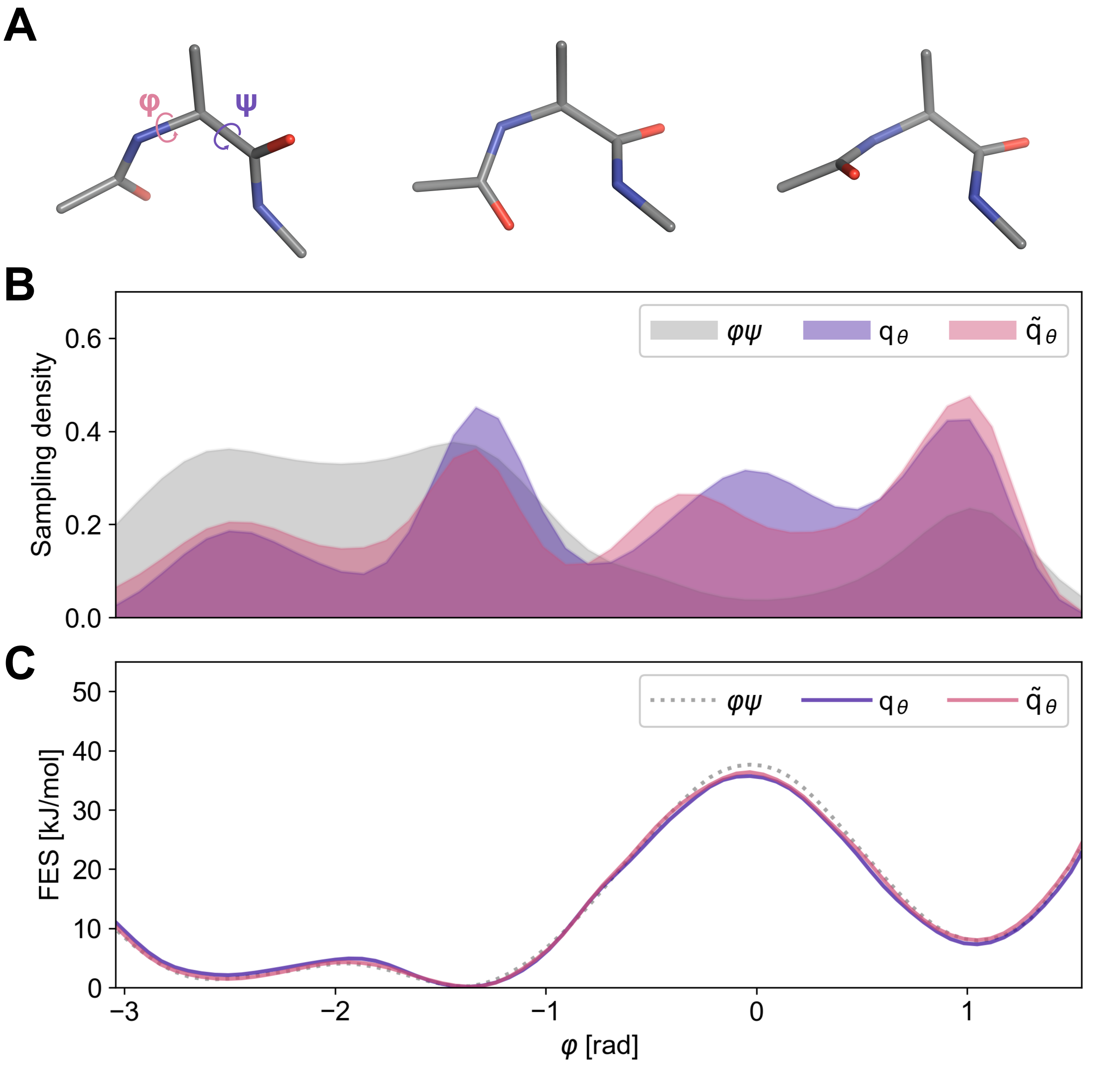}
        \caption{\textbf{Alanine dipeptide isomerization.} \textbf{A.} Representative snapshots of the isomerization process.
        \textbf{B.} Sampling density distribution projected along the $\phi$ torsional angle.
        \textbf{C.} Free energy surface (FES) plotted along the $\phi$ torsional angle. 
        In panels \textbf{B} and \textbf{C}, results obtained with a standard OPES run using $\phi$ and $\psi$ torsional angles as CVs are reported in grey, those obtained with OPES+V$_K$ based on a committor model trained following the original variational approach (q$_{\boldsymbol{\theta}}$) are reported in purple, and those obtained with OPES+V$_K$ based on a committor model trained following the simplified variational approach ($\tilde{\text{q}}_{\boldsymbol{\theta}}$) are reported in pink.}
        \label{fig:alanine}
    \end{figure}

This loss is complemented by the same boundary term $L_b$ used in the original recipe, which is expressed as
    \begin{equation}
        L_b =
        \frac{1}{N_A} \sum_{i \in A} q(\mathbf{x}_i)^2
        +
        \frac{1}{N_B} \sum_{i \in B} \left(q(\mathbf{x}_i) - 1\right)^2
        \label{eq:lbound}
    \end{equation}
and is evaluated on a labeled dataset of $N_A$ and $N_B$ configurations collected by running short, unbiased MD simulations in the two basins. 
The overall optimization of the network parameters is then performed by minimizing a combination of the two losses 
$L = L_v + \alpha L_b$, where the $\alpha$ hyperparameter scales the relative magnitude of the two contributions. 

As a final practical note, we recall that, for improved numerical stability, it is also possible to minimize $\log(L)$, as proposed in Ref.~\citenum{deng2026fluctuations}.
In addition, we found that the new learning objective is more prone to overfitting in some cases.
It is thus advisable to monitor this behavior and, if necessary, adopt strategies to avoid it, e.g., via early stopping or shorter training.

\section{Results}
\subsection{Alanine Dipeptide conformational equilibrium}
    The first test system for our methodology was the conformational equilibrium of alanine dipeptide, often used as a benchmark for new enhanced sampling methods (see Fig.~\ref{fig:alanine}A). 
    As often done in the past for this molecule, we use the interatomic distances between the heavy atoms of the molecule as input descriptors.~\cite{bonati2020data, bonati2021deep, trizio2021enhanced, trizio2025everything}
    Based on this description of the system, we set up our simplified approach, obtaining an approximated committor function, which was used to drive enhanced sampling calculations with the OPES and Kolmogorov biases.
    This provided effective sampling of the entire relevant phase space by promoting frequent transitions between the metastable states (see Fig.~\ref{fig:alanine}B).
    From the data thus collected, we recovered accurate free energy estimates that were almost indistinguishable from those obtained using as CVs the conventional $\phi$-$\psi$ torsional angles or a committor model trained on the original variational approach (see Fig.~\ref{fig:alanine}C).    
    
    Notably, our calculations also retained a strong focus on the TS region, whose sampling is greatly enhanced if compared with a standard OPES run (see Fig.~\ref{fig:alanine}B).
    Nonetheless, the sampling density is not exactly centered on the \textit{true} transition state, which is located at $\phi \sim 0$.
    Even if in terms of sampling, this is only a minor sacrifice, for completeness, one may be interested in obtaining a more accurate estimate of the committor for further analysis.
    To this end, it is possible to use the cheaply generated data from the simplified approach to perform a single additional iteration using the original, more expensive variational approach, without the need for multiple calculations.
    As expected, the results obtained with the original strategy are slightly better in terms of TS sampling, but remarkably, no differences can be seen in the FES estimate. 

    However, it must be noted that, even on a system as simple as alanine and with a limited set of simple descriptors, the training time using the original formulation was roughly 3 times higher than using the simplified one, thus motivating the use of the latter, especially for more complex systems.

\subsection{Proton transfer in tropolone}
\begin{figure}[h!]
        \centering
        \includegraphics[width=1\linewidth]{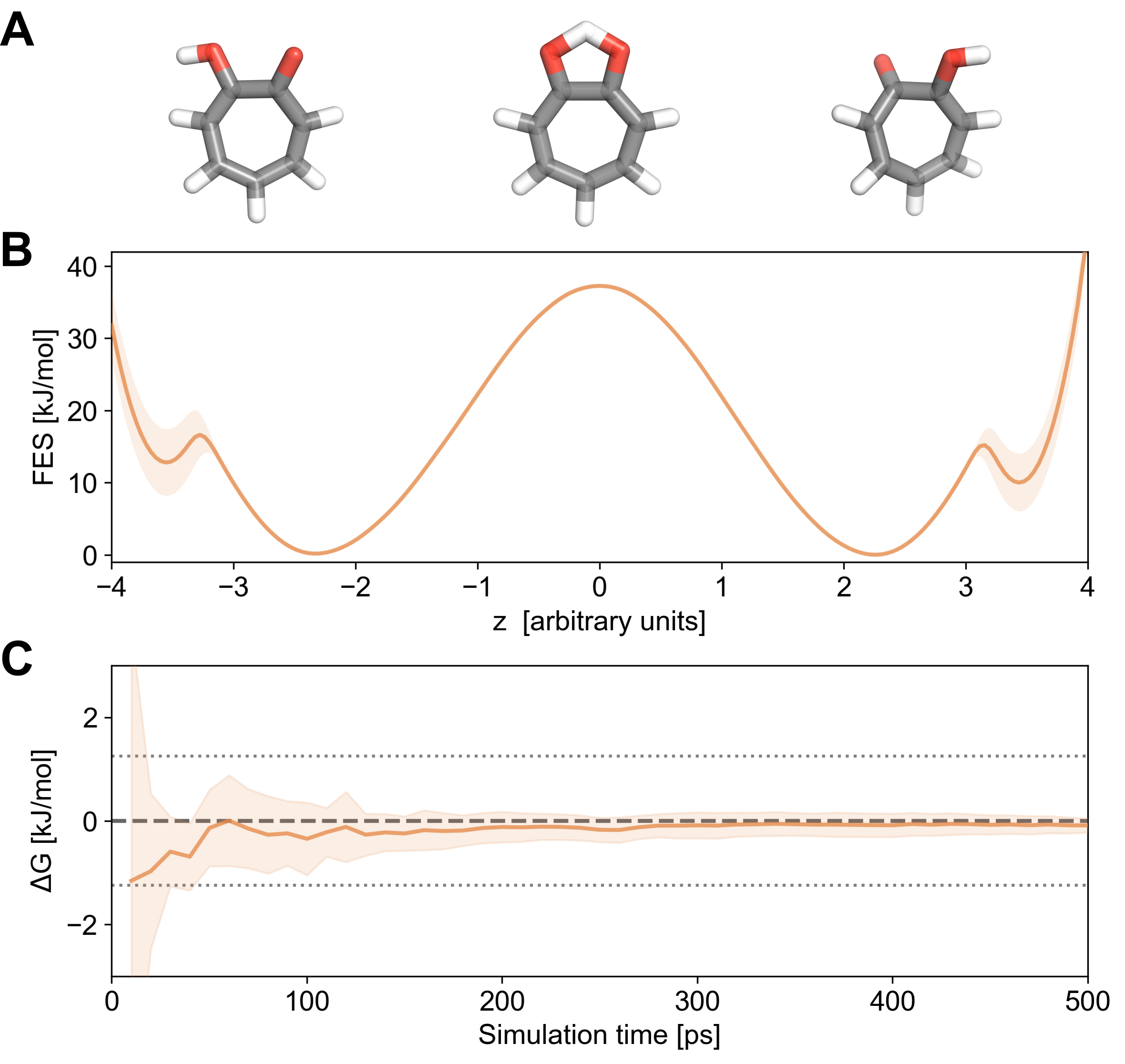}
        \caption{\textbf{Proton transfer in tropolone.} \textbf{A.} Representative snapshots of the reaction stages, with the transition state characterized by the shared proton between the two oxygens.
        \textbf{B.} Free energy surface (FES) plotted along the learned $z$ CV. 
        \textbf{C.} Convergence with simulation time of the estimated free energy difference between the two main isomers. The dashed line shows the reference value, with the $\pm 0.5 \,$k$_B$T interval marked by the thin dotted lines. 
        In panels \textbf{B} and \textbf{C}, average values obtained from 3 independent simulations are depicted as solid lines, whereas the corresponding standard deviations are shown as shaded areas. 
        }
        \label{fig:tropolone}
    \end{figure}

    As a second test case, we studied the intramolecular proton transfer (PT) involved in the isomerization of tropolone, a small aromatic molecule.
    The aromatic ring of this molecule is substituted with two adjacent oxygens, which, depending on the specific isomer, can be found either in the hydroxyl or keto form (see Fig.~\ref{fig:tropolone}A).
    In addition, two other minor metastable states are present, determined by the relative orientation (inward or outward) of the H of the hydroxyl group with respect to the keto oxygen.

    Here, the input descriptors used were the interatomic distances involving the heavy atoms and those between the reactive H and O atoms.
    The four metastable states mentioned above are clearly distinguishable in the FES estimate obtained with our method (see Fig.~\ref{fig:tropolone}B).
    As expected, the FES is also symmetric, as the two oxygens are equivalent, and the free energy difference between the two isomers quickly converges to the correct value $\Delta G = 0\,$kJ/mol (see Fig.~\ref{fig:tropolone}C). 

\subsection{OAMe-G2 binding}
    If the first two examples could have been handled with the original approach at only a moderate additional computational cost, we now move to systems where that would have been more problematic.
    One such case is the ligand-binding process in solution, in which it is necessary to account for the solvent role in the dynamics of both the binding pocket and the ligand itself.
    Unfortunately, as the solvent molecules are indistinguishable and highly mobile, applying the original approach requires an explicit dependence of the model on a large number of atoms.
    At best, this limitation introduces several practical issues in handling and processing data related to many atoms, which can easily saturate available computing memory and dramatically slow down training.
    Indeed, even in simplified systems with a limited number of atoms and solvent molecules, as the one discussed here, the original approach can be orders of magnitude slower than the simplified one proposed here.
    In even more complex scenarios, the limitations of the original framework make its application unfeasible in practice.

\begin{figure}
        \centering
        \includegraphics[width=1\linewidth]{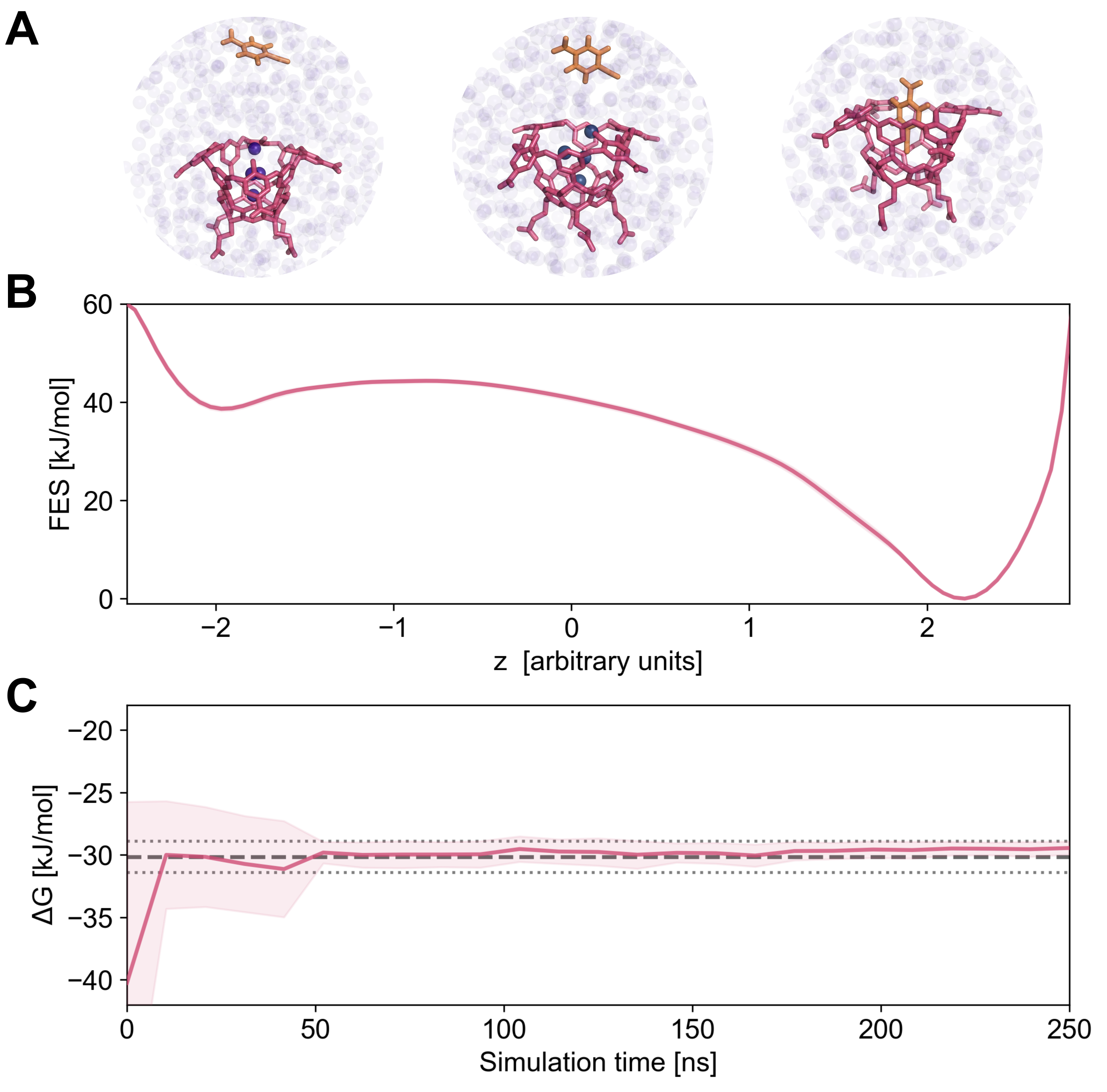}
        \caption{\textbf{G2-OAMe binding.} 
        \textbf{A.} Representative snapshots of the binding stages, with the host molecule colored in pink and the guest in orange. Water molecules are represented as blue spheres in solid color when inside the pocket, and transparent otherwise.
        \textbf{B.} Free energy surface (FES) plotted along the learned $z$ CV. 
        \textbf{C.} Convergence with simulation time of the estimated binding energy. The dashed line shows the reference value obtained with the same setup as Ref.~\citenum{rizzi2021role}, with the $\pm 0.5 \,$k$_B$T interval marked by the thin dotted lines. 
        In panels \textbf{B} and \textbf{C}, average values obtained from 3 independent simulations are depicted as solid lines, whereas the corresponding standard deviations are shown as shaded areas.}
        \label{fig:calixarene}
    \end{figure}

    As a prototypical example of this class of processes, we studied the binding of the G2 molecule to the OAMe calixarene guest from the SAMPL5 challenge,~\cite{yin2017overview} which has been extensively characterized in several papers~\cite{bhakat2017resolving,rizzi2021role,ray2023deep,trizio2025everything} (see Fig.~\ref{fig:calixarene}A).
    As done in Refs.~\citenum{rizzi2021role} and \citenum{trizio2025everything}, we chose to treat the solvation degrees of freedom separately from the trivial positional ones.
    To this end, we trained our model solely on water coordination numbers, computed for a few ligand atoms and a set of virtual atoms above the binding pocket, and biased the resulting committor-like CV using both OPES and $V_K(\textbf{x})$. 
    Specifically, the OPES component of the bias was applied to both the committor-based CV $z(\textbf{x})$ and the ligand-pocket distance.    
    Following this strategy, we were able to promote efficient sampling and obtain accurate free energy estimates in good agreement with previous studies with a similar computational setup (see Fig.~\ref{fig:calixarene}B and C).

    It must be noted that, even with the relatively simple setup and the limited scale of this system, in order to use the old formulation, one would have needed to track, and thus print to file, the positions of a few hundred water molecules per frame.
    In contrast, with the approach proposed in this work, the required data is reduced to the simple set of 12 coordination numbers used as descriptors, whose storage is negligible even for long trajectories. 
    Most remarkably, the new scheme led to a 100-fold reduction in the training computational burden while still retaining a more than satisfactory sampling.

    \begin{figure}[t]
        \centering
        \includegraphics[width=1\linewidth]{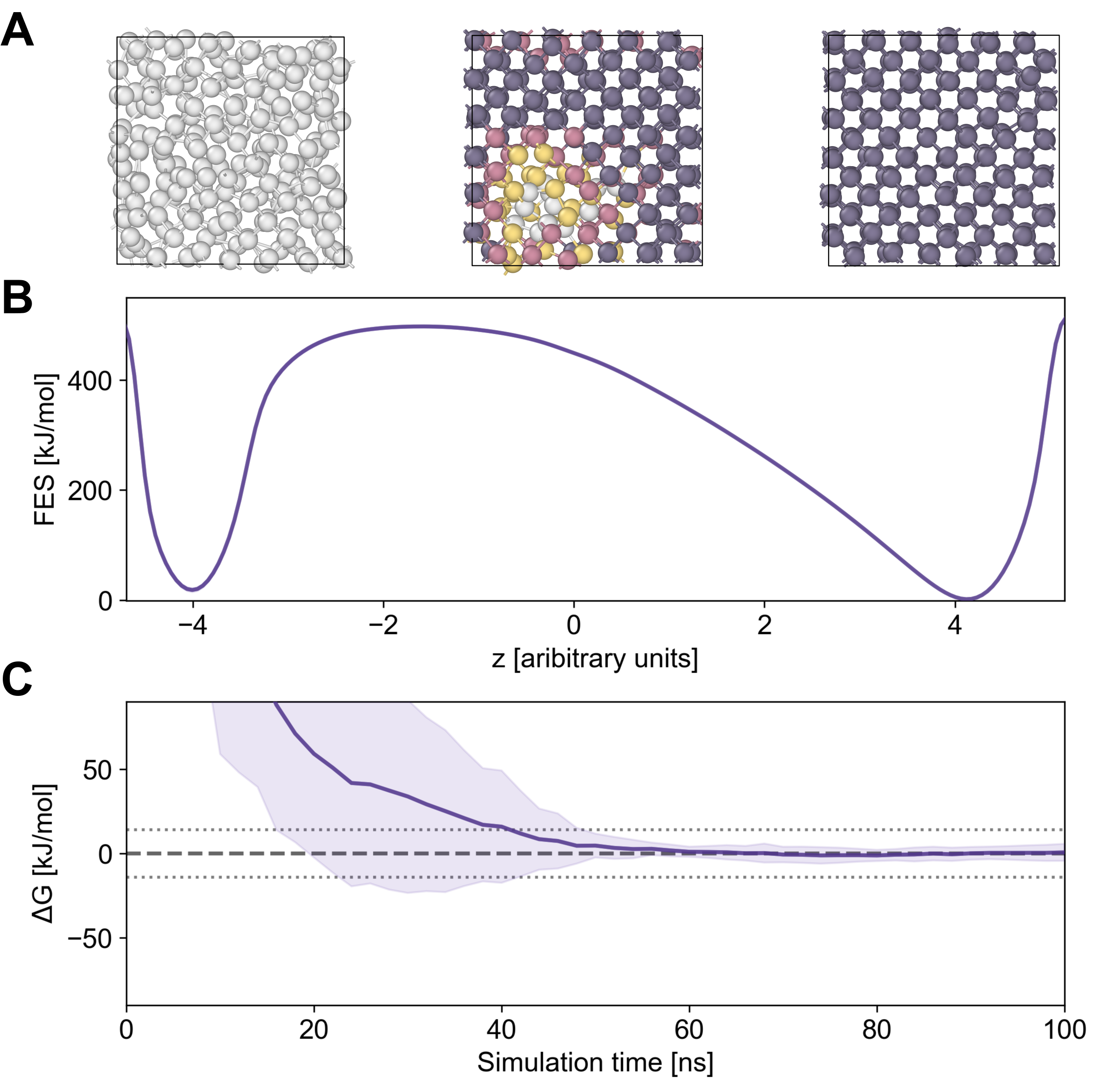}
        \caption{ \textbf{Silicon crystallization.} 
        \textbf{A.} Representative top-view snapshots of the crystallization process from the disordered liquid phase to the ordered solid.
        Atoms are colored according to the degree of similarity of their local arrangement to the ideal diamond structure. The darker the color, the higher the similarity. 
        \textbf{B.} Free energy surface (FES) plotted along the learned $z$ CV. 
        \textbf{C.} Convergence with simulation time of the estimated free energy difference between the two states. The dashed line shows the reference value, with the $\pm\,$k$_B$T interval marked by the thin dotted lines. 
        In panels \textbf{B} and \textbf{C}, average values obtained from 3 independent simulations are depicted as solid lines, whereas the corresponding standard deviations are shown as shaded areas.}
        \label{fig:silicon}
    \end{figure}

\subsection{Crystallization of Si}
    As a final test case, we studied the crystallization of silicon to its diamond structure close to the melting temperature (T=1700K), a scenario where the practical advantages of our new approach are most evident (see Fig.~\ref{fig:silicon}A).
    In fact, when studying phase transitions, it is necessary to properly account for symmetries and local atomic arrangements, which typically requires descriptors that are complex, many-body functions of the atomic coordinates, such as the Steinhardt order parameters, and possibly depend on \textit{all} the particles in the simulation box, for example, radial density functions or x-ray diffraction peak intensities.~\cite{niu2018silica}
    This would clearly represent a significant bottleneck within the original framework, but becomes negligible with our new formulation.  

    Indeed, following Ref.~\citenum{bonati2021deep}, we employed as input descriptors the peaks of the anisotropic structure factor of silicon.
    In this case, considering the different relative magnitudes of these inputs, we also introduced a preprocessing normalization layer to the NN architecture. 
    In addition, to make up for the expected larger free energy barrier and the considerably high temperature, we used a larger $\lambda$ value, i.e., $\lambda=1000$.

    Even starting from data limited to the pure phases, our iterative procedure allowed us to progressively learn an effective reaction coordinate, which correlates well with the commonly used, but rather expensive, CVs based on environment similarity~\cite{piaggi2019phase} and Steinhardt parameters (see supplementary material, Fig.~\ref{SI_fig:silicon_comparison}).
    
    Remarkably, also for this more complex process, many reactive events were observed.
    From such data, it was possible to recover accurate free energy estimates (see Fig.~\ref{fig:silicon}B), with the two phases found at the same relative free energy, as expected when being at the melting temperature (see Fig.~\ref{fig:silicon}C).

\section{Discussion}

    This paper fits into a broader effort to extend enhanced sampling methods to the detailed study of transition states and to make them more automated and accessible. 
    In Refs.~\citenum{kang2024computing} and \citenum{trizio2025everything}, we leveraged machine learning together with the longstanding formal theory of the committor function to achieve first extensive sampling of the TS region, and later uniform and ergodic sampling of the relevant phase space. 
    With Ref.~\citenum{kang2026without}, we moved a step further toward automation by exploiting the expressive power of graph neural networks to operate directly on atomic coordinates, thereby bypassing the need for manually designed input descriptors, although at a higher computational cost.
    
    In this work, we focus on accessibility and scalability, introducing a controlled approximation of the committor function to dramatically reduce the practical barrier to use. 
    By reframing the learning procedure as a simpler variational problem, we reduce the optimization complexity and its associated computational cost, achieving orders-of-magnitude savings in training time. 
    This reduction is particularly impactful in more complex systems, making the approach feasible when it would otherwise be prohibitive. 
    At the same time, even for simpler cases, the approximated formulation yields a more streamlined workflow, significantly simplifying the setup and accelerating calculations, making it naturally suitable for automated, scalable computational pipelines.
    
    These advantages, however, come with a trade-off that should be acknowledged, as the new formulation does not formally target the \textit{true} committor function, a defining feature of the original approach and central to its use in detailed mechanistic analysis. 
    Nonetheless, when no other viable option is available, the approximated committor can still provide valuable insight as well as clear sampling advantages.
    In particular, an estimate of the original variational functional, informative of kinetic rates, can be recovered from the approximated model with a single calculation and without any additional optimization, leveraging Eq.~\ref{eq:functional_split}.
    
    From a broader perspective, the proposed framework can also serve as an entry point within multistage workflows. 
    It can be deployed as a fast and robust sampling engine during early exploration of complex or computationally intensive systems, providing an initial characterization at a fraction of the cost.
    Subsequently, the formally exact but more demanding approach can be adopted for final production simulations and in-depth analysis.
    
    Overall, this work represents an important step toward making committor-based enhanced sampling practical at scale. 
    By significantly lowering the computational cost while mostly retaining the original framework’s sampling power, the proposed approach extends the reach of TS-focused simulations and contributes to consolidating committor-based methods as a scalable tool within the enhanced sampling toolbox for the study of complex systems.

\section*{Supplementary Material}
    The supplementary material includes the extended derivation of the simplified variational functional proposed here, additional figures, and the computational details of the simulations performed and training of the approximated committor models.

\section*{Code and Data Availability} \label{sec:code_avail}
    The committor models have been trained using the open-source library \texttt{mlcolvar}.~\cite{bonati2023mlcolvar}
    Enhanced sampling simulations have been performed using a custom interface for the PLUMED~\cite{plumed2019promoting} plugin for enhanced sampling and free energy calculations.
    Simulation inputs and the code needed for training and performing enhanced sampling simulations will be made available on GitHub upon publication.
        
\begin{acknowledgments}
    The authors are grateful to Luigi Bonati, Florian Dietrich, Ioannis Galdadas, Peilin Kang and Alice Triveri for useful feedback and discussions on the manuscript.
\end{acknowledgments}

\section*{Competing interests statement}
    The authors declare no competing interests.

\section*{References}
\putbib

\end{bibunit}

\clearpage
\onecolumngrid
\begin{bibunit}

\setcounter{page}{1}
\renewcommand{\thepage}{S\arabic{page}}
\setcounter{section}{0}
\renewcommand{\thesection}{S\arabic{section}}
\setcounter{equation}{0}
\renewcommand{\theequation}{S\arabic{equation}}
\setcounter{figure}{0}
\renewcommand{\thefigure}{S\arabic{figure}}
\setcounter{table}{0}
\renewcommand{\thetable}{S\arabic{table}}
    
\clearpage

{\Large\normalfont\sffamily\bfseries{{Supporting Information}}}

\setlength{\tabcolsep}{18pt}
\renewcommand{\arraystretch}{1.2}

\setlength{\abovecaptionskip}{0.5pt} 

\section{Reformulation of Kolmogorov functional}

Here, we outline the steps leading to the simplified variational principle proposed here.
For compactness, we drop the dependence on atomic coordinates and descriptors, i.e., $q(\textbf{d}(\textbf{x})) \rightarrow q $.

To compute $K[q]=\langle |\nabla q|^2\rangle$, we sum over the 3 spatial dimensions $l$ and the $N$ atoms $n$.
\begin{equation}
    K[q] =
    \langle |\nabla q|^2\rangle =
    \bigg\langle \sum_l^3 \sum_n^N \left( \frac{\partial q}{\partial u_{l,n}}\right)^2 \bigg\rangle 
\end{equation}
where $u$ denotes the mass-scaled Cartesian coordinates.
We can then re-write the derivatives in the last term using the chain rule. We first take the derivatives of $q$ with respect to descriptor $i$, and then of descriptor $i$ with respect to atom $n$. The resulting expression is written as a sum over the $N_d$ descriptors (index $i$), in addition to the sums over the atoms $n$ and the Cartesian components $l$.
\begin{equation}
    \bigg\langle \sum_l^3 \sum_n^N \left( \frac{\partial q}{\partial u_{l,n}}\right)^2 \bigg\rangle =
    \bigg\langle \sum_l^3 \sum_n^N \left( \sum_i^{N_d} \frac{\partial d_i}{\partial u_{l, n}}\frac{\partial q}{\partial d_i}\right)^2 \bigg\rangle 
\end{equation}

We can now expand the object in the parentheses. To keep track of the cross elements, we need to split the descriptor indices, i.e., $i \rightarrow i,j$

\begin{equation}
    \bigg\langle \sum_l^3 \sum_n^N \left( \sum_i^{N_d} \frac{\partial d_i}{\partial u_{l, n}}\frac{\partial q}{\partial d_i}\right)^2 \bigg\rangle  =
    \bigg\langle \sum_l^3 \sum_n^N \sum_{i,j}^{N_d}
    \frac{\partial d_i}{\partial u_{l, n}}\frac{\partial q}{\partial d_i}
    \frac{\partial d_j}{\partial u_{l, n}}\frac{\partial q}{\partial d_j}
    \bigg\rangle
\end{equation}

Since the derivatives of $q$ with respect to the descriptors do not depend on atom indices, the coordinate-dependent terms can be grouped separately. 
The sums over atoms and spatial components can thus be collected into a purely geometric factor that depends only on the descriptor gradients with respect to the atomic coordinates. The remaining factor depends exclusively on the derivatives of $q$ in descriptor space.

\begin{equation}
    \bigg\langle \sum_l^3 \sum_n^N\sum_{i,j}^{N_d} 
    \frac{\partial d_i}{\partial u_{l, n}}\frac{\partial q}{\partial d_i}
    \frac{\partial d_j}{\partial u_{l, n}}\frac{\partial q}{\partial d_j}
    \bigg\rangle = \bigg\langle \sum_{i,j}^{N_d}
    \left(\sum_l^3 \sum_n^N \frac{\partial d_i}{\partial u_{l, n}} \frac{\partial d_j}{\partial u_{l, n}} \right)
    \frac{\partial q}{\partial d_i} \frac{\partial q}{\partial d_j}
    \bigg\rangle
\end{equation}

This way, the functional can be decoupled in two contributions: a geometric term determined by the descriptor set and a model-dependent one, constructed from the descriptor derivatives of $q$. 

\begin{equation}
    K[q]  =
    \bigg\langle \sum_{i,j}^{N_d}
    \underset{g_{ij}}{\underline{\left( \sum_l^3 \sum_n^N \frac{\partial d_i}{\partial u_{l, n}} \frac{\partial d_j}{\partial u_{l, n}} \right) }}
    \underset{k_{ij}}{\underline{\frac{\partial q}{\partial d_i} \frac{\partial q}{\partial d_j}}} \bigg\rangle = 
    \langle \sum_{i,j}^{N_d} g_{ij} k_{ij} \rangle \overset{\text{implicit}}{\rightarrow} 
    \langle g_{ij} k_{ij} \rangle
\end{equation}
In the last step, we adopt Einstein summation convention over the descriptor indices $i$ and $j$. For each configuration in the ensemble,  $\textbf{G}$ and $\textbf{K}$ are symmetric matrices of size $N_d\times N_d$.
In this form, the functional can be written as the ensemble average of the contraction between the geometric and model-dependent terms,
\begin{equation}
    \langle |\nabla q|^2\rangle = \langle g_{ij} k_{ij} \rangle
\end{equation}

We then apply the \textit{Cauchy-Schwarz inequality} to decouple the contributions and motivate our approach of skipping the positions dependence, which is now encoded only in $g_{ij}$.
This leads to the inequality reported in the main text (see Eq.~\ref{eq:cauchy_inequality})
\begin{equation}
     \langle g_{ij} k_{ij} \rangle^2 \leq 
     \langle g_{ij} g_{ji} \rangle 
     \langle k_{ij} k_{ji} \rangle 
\end{equation}
which motivates treating the geometric dependent term separately and focusing on the descriptor-space contribution in the practical implementation.

\section{Computation of \textbf{K} and \textbf{G} matrices}
Instead of computing the quantity
    \begin{equation}
        k_{ij} k_{ji} = \operatorname{tr}(\textbf{K}^2)
    \end{equation}
via explicit matrix multiplication, it is computationally convenient to rewrite it using some elementary algebra.

For completeness, we also report similar manipulations for the $\textbf{G}$ matrix for the computation of
    \begin{equation}
        g_{ij} g_{ji} = \operatorname{tr}(\textbf{G}^2)
    \end{equation}

\subsection{Trace identity for \textbf{K}}
Matrix $\textbf{K}$ is defined component-wise as
    \begin{equation}
        k_{ij} = h_i h_j = \frac{\partial q}{\partial d_i}\frac{\partial q}{\partial d_j} \qquad \text{with} \qquad \textbf{h} =
        \begin{pmatrix}
        \frac{\partial q}{\partial d_1} \\\vdots \\
        \frac{\partial q}{\partial d_{N_d}}
        \end{pmatrix}
    \end{equation}
In matrix form, this can be written as the outer product $\textbf{K} = \textbf{h} \textbf{h}^\top$. Squaring the matrix gives
    \begin{equation}
        \textbf{K}^2 = \textbf{h h}^\top \textbf{h} \textbf{h}^\top = (\textbf{h}^\top \textbf{h})\, \textbf{h h}^\top
    \end{equation}
Taking the trace, we obtain
    \begin{equation}
        \operatorname{tr}(\textbf{K}^2)
        = \operatorname{tr}\!\left( (\textbf{h}^\top \textbf{h})\, \textbf{h h}^\top \right)
        = (\textbf{h}^\top \textbf{h})\, \operatorname{tr}(\textbf{h h}^\top)
    \end{equation}
Since $\operatorname{tr}(\textbf{h h}^\top) = \textbf{h}^\top \textbf{h}$, it follows that
    \begin{equation}
        \operatorname{tr}(\textbf{K}^2) = (\textbf{h}^\top \textbf{h})^2 = \|\textbf{h}\|^4
        \label{eq:trace_k2_simplified}
    \end{equation}

\subsection{Trace identity for \textbf{G}}
For a generic matrix $\textbf{G}$, the Frobenius norm is defined as
    \begin{equation}
        \|\textbf{G}\|_F^2 = \sum_{i}\sum_{j} g_{ij}^2 
    \end{equation}
On the other hand, using the definition of matrix multiplication,
    \begin{equation}
        (\textbf{G}^\top \textbf{G})_{ii} = \sum_{j} g^\top_{ij} g_{ji}
        = \sum_{j} g_{ji} g_{ji} 
    \end{equation}
Taking the trace of $\textbf{G}^\top \textbf{G}$ gives 
    \begin{equation}
        \operatorname{tr}(\textbf{G}^\top \textbf{G})
        = \sum_{i} (\textbf{G}^\top \textbf{G})_{ii}
        = \sum_{i}\sum_{j} g_{ji}^2
        = \sum_{i}\sum_{j} g_{ij}^2 
    \end{equation}
Therefore, we can write 
    \begin{equation}
        \operatorname{tr}(\textbf{G}^\top \textbf{G}) = \|\textbf{G}\|_F^2  
    \end{equation}
If matrix $\textbf{G}$ is symmetric (\(\textbf{G}^\top=\textbf{G}\)), then this simplifies to
    \begin{equation}\label{eq:frobenius}
        \operatorname{tr}(\textbf{G}^2)=  \|\textbf{G}\|_F^2 
    \end{equation}
We now define the matrix \textbf{G} as in Eq.~\ref{eq:gk_terms}
    \begin{equation}
        g_{ij}
        =
        \sum_{l}^{3}
        \sum_{n}^{N}
        \frac{\partial d_i}{\partial u_{l, n}}
        \frac{\partial d_j}{\partial u_{l, n}} 
    \end{equation}
Introducing
    \begin{equation}
        J_{i\alpha} = \frac{\partial d_i}{\partial u_\alpha},
        \qquad \alpha = (n,l), \quad \alpha = 1,\dots,3N
    \end{equation}
the matrix can be written as
\begin{equation}
    g_{ij} = \sum_{\alpha}^{3N} J_{i\alpha}\, J_{j\alpha} 
\end{equation}
Equivalently, in matrix form,
    \begin{equation}
        \textbf{G} = \textbf{J} \textbf{J}^\top
    \end{equation}
\medskip
Using Eq.\ref{eq:frobenius}, since $\textbf{G}$ is symmetric, gives
    \begin{equation}
        \operatorname{tr}(\textbf{G}^2) = \|\textbf{G}\|_F^2 =  \| \textbf{J} \textbf{J}^\top\|_F^2 
    \end{equation}

\clearpage
\section{Alanine Dipeptide conformational equilibrium}
\subsection{Computational details}
        \paragraphtitle{Simulations details}
            The alanine dipeptide (Ace-Ala-Nme) simulations in vacuum were carried out targeting the NVT ensemble using the GROMACS-2024.5~\cite{abraham2015gromacs} MD engine patched with PLUMED-2.9.2~\cite{tribello2014plumed, plumed2019promoting} and the Amber99-SB~\cite{amber2013} force field with a 2~fs timestep. 
            The Langevin dynamics is sampled with damping coefficient $\gamma_i= \frac {m_i}{\tau-t}$ with $\tau-t = 0.05$~ps and target temperature of 300~K.

    \paragraphtitle{Approximated committor model training details}
        To model the committor function $q_\theta(\textbf{x})$ at each iteration, we used the 45 distances between all heavy atoms as inputs of a neural network (NN) with architecture  [45, 32, 32, 1] nodes/layer. 
        For the optimization, we used the ADAM optimizer with an initial learning rate of $10^{-3}$ modulated by an exponential decay with multiplicative factor $\gamma=0.99995$. 
        The training was performed for 5000 epochs. 
        The $\alpha$ hyperparameter in the loss function was set to 10$^3$, and we optimized the $\log$ of the total loss for numerical stability. 
        The number of iterations, the corresponding dataset size, and the $\lambda$ and the OPES \texttt{BARRIER} used in the biased simulations with the corresponding model are summarized in Table~\ref{SI_tab:alanine_iterations} alongside the simulation time $t_s$ and the output sampling time $t_o$.
            \begin {table}[h!]
                \caption {\textbf{Summary alanine.} Summary of the iterative procedure for alanine.} \label{SI_tab:alanine_iterations}
                \begin{center}
                \begin{tabular}{ |c|c|c|c|c|c| } 
                 \hline
                 Iteration & Dataset size & OPES \texttt{BARRIER} [kJ/mol] & $\lambda$ & $t_s$ [ns] & $t_o$ [ps] \\ 
                 \hline
                    0   & 5000 & 30 & 2   & 2*10 & 1 \\
                    1   & 22000 & 30 & 2 & 2*10 & 1 \\
                    2   & 40000 & 30 & 3 & 2*10 & 1 \\ 
                 \hline
                \end{tabular}
                \end{center}
            \end {table}

    \subsection{Additional figures}
    \begin{figure}[h!]
        \centering
        \includegraphics[width=\linewidth]{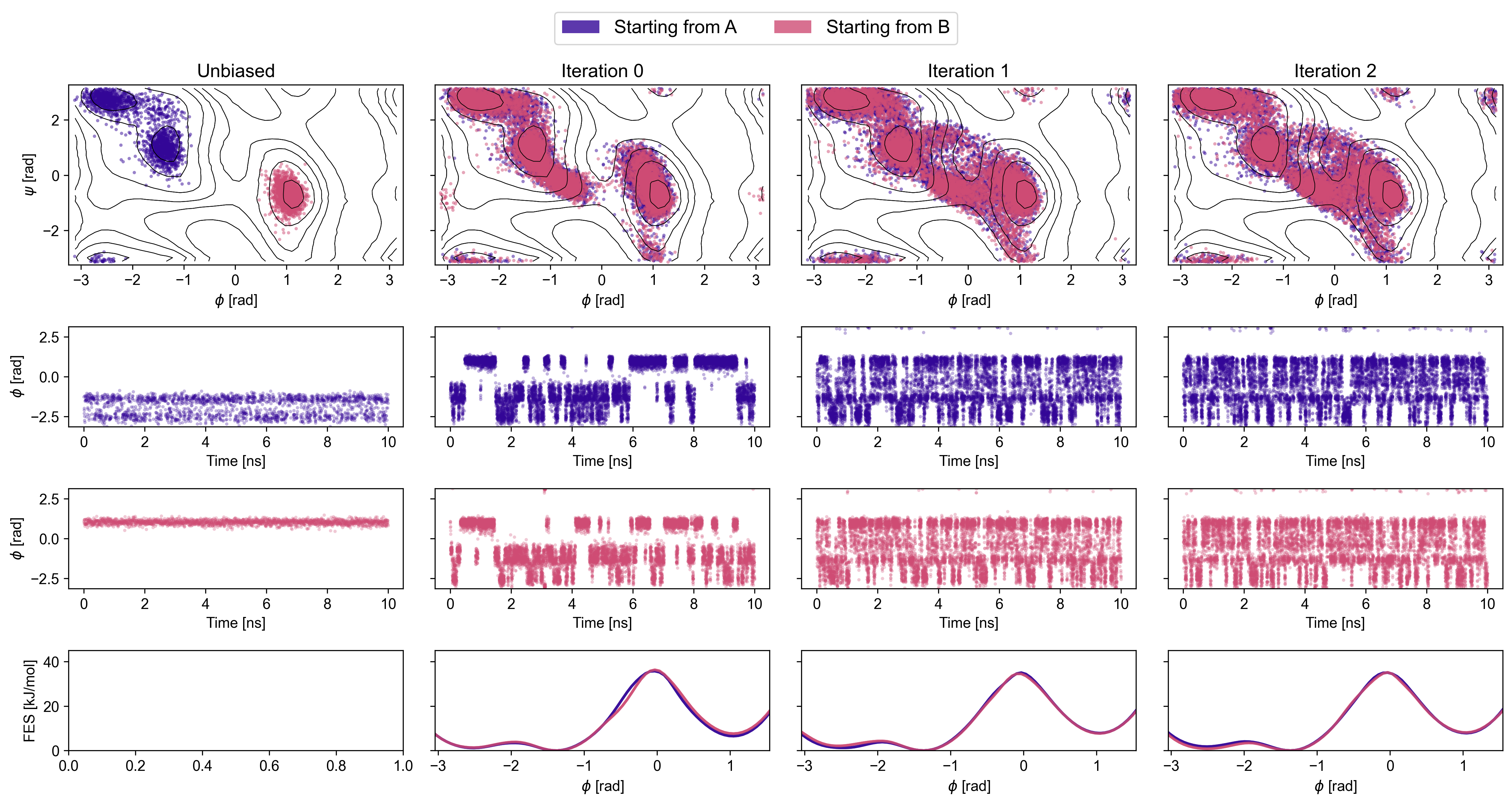}
        \caption{\textbf{Iterative sampling on alanine.} Top row:  Scatter plot of the sampled points in the $\phi\psi$ space at successive iterations. Points from trajectories initialized in state A are shown in purple, while those initialized in state B are shown in pink.
        Second row: Time series of the $\phi$ dihedral angle for trajectories started from state A.
        Third row: Time series of the $\phi$ dihedral angle for trajectories started from state B.
        Bottom row: Free energy surface (FES) plotted along $\phi$, estimated from the sampled configurations at each iteration.}
        \label{SI_fig:alanine_sampling}
    \end{figure}
    
    \begin{figure}[h!]
        \centering
        \includegraphics[width=\linewidth]{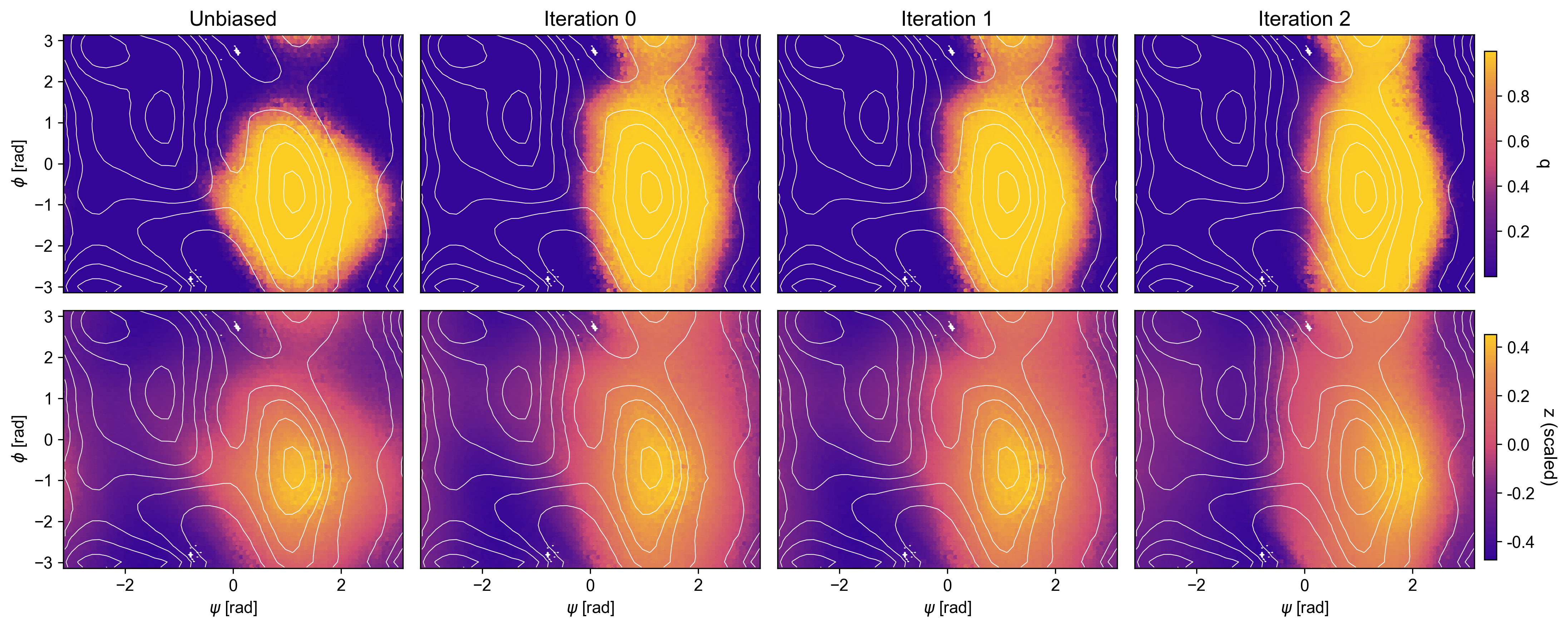}
        \caption{\textbf{Alanine committor models.} Contour plots in the $\phi\psi$ space of the learned $q$ (top row) model and the corresponding $z$ (bottom row) at successive iterations. Values of the $z$ variable are scaled to be in the [-0.5, 0.5] range.}
        \label{SI_fig:alanine_models}
    \end{figure}
    
    \begin{figure}[h!]
        \centering
        \includegraphics[width=\linewidth]{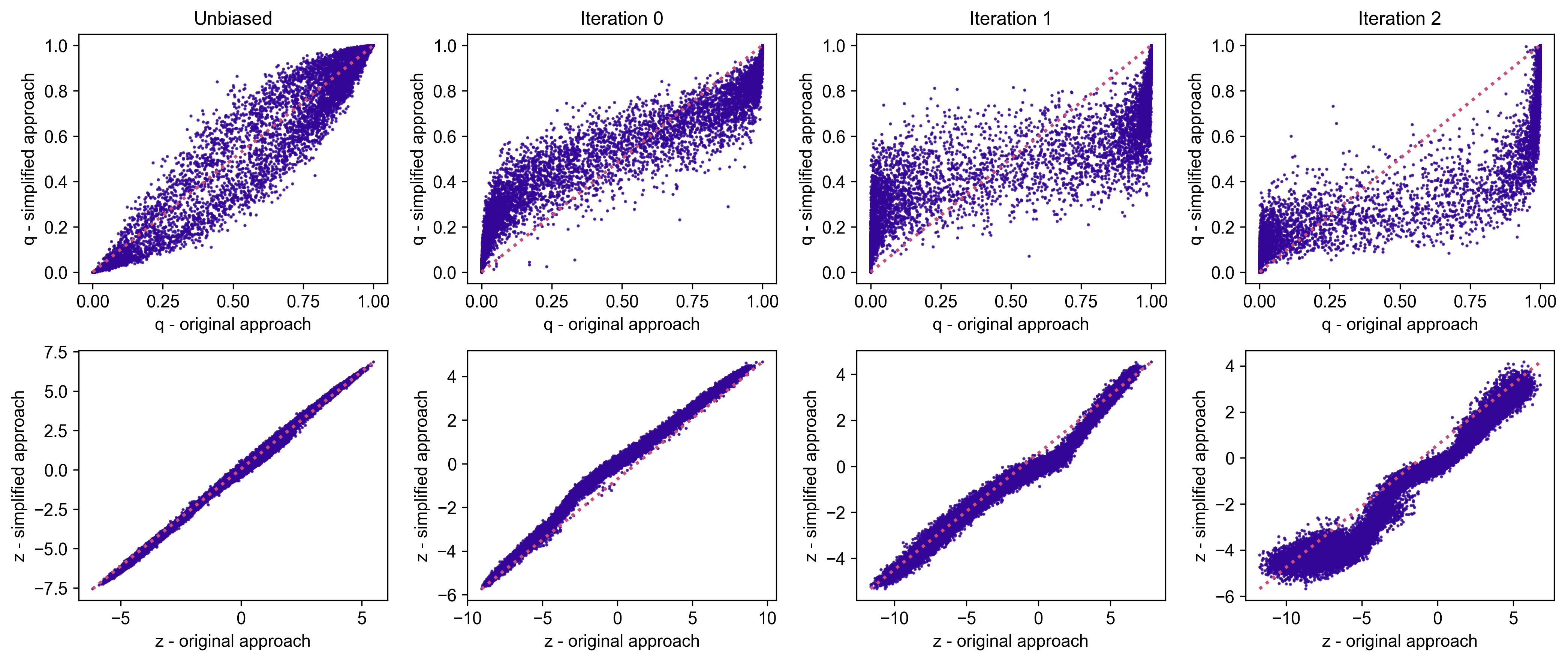}
        \caption{\textbf{Convergence of alanine committor models.} Comparison between the $q$ (top row) and the corresponding $z$ learned using the original and the simplified approaches at successive iterations.}
        \label{S_fig:alanine_comparison}
    \end{figure}

\clearpage
\section{Tropolone proton transfer}
    \subsection{Computational details}
        \paragraphtitle{Simulations details}
        The intramolecular proton transfer simulations were carried out targeting the NVT ensemble using the CP2K-8.1~\cite{kuhne2020cp2k} software package patched with PLUMED-2.9.1~\cite{plumed2019promoting} at PM6 semi-empirical level. 
        The integration step was 0.5~fs and we used the velocity rescaling thermostat~\cite{bussi2007velocity} set at 300~K with a time constant of 100~fs.
        
        \paragraphtitle{Approximated committor model training details}
        To model the committor function $q_\theta(\textbf{x})$ at each iteration, we used the 36 distances between all heavy atoms of the molecule and the distances between the reactive hydrogen and the two oxygens as inputs of a neural network (NN) with architecture  [38, 16, 16, 1] nodes/layer. 
        For the optimization, we used the ADAM optimizer with an initial learning rate of $10^{-3}$ modulated by an exponential decay with multiplicative factor $\gamma=0.99995$. 
        The training was performed for 5000 epochs in the first cycle and then 20000 epochs in the following ones. 
        The $\alpha$ hyperparameter in the loss function was set to 10$^3$, and we optimized the $\log$ of the total loss for numerical stability. 
        The number of iterations, the corresponding dataset size, and the $\lambda$ and the OPES \texttt{BARRIER} used in the biased simulations with the corresponding model are summarized in Table~\ref{SI_tab:tropolone_iterations} alongside the simulation time $t_s$ and the output sampling time $t_o$.
            \begin {table}[h!]
                \caption {\textbf{Summary tropolone PT.} Summary of the iterative procedure for intramolecular proton transfer in tropolone.} \label{SI_tab:tropolone_iterations}
                \begin{center}
                \begin{tabular}{ |c|c|c|c|c|c| } 
                 \hline
                 Iteration & Dataset size & OPES \texttt{BARRIER} [kJ/mol] & $\lambda$ & $t_s$ [ps] & $t_o$ [ps] \\ 
                 \hline
                    0   & 4000 & 60 & 1.2   & 2*500 & 0.05 \\
                    1   & 22000 & 60 & 4  & 2*500 & 0.05 \\
                    2   & 40000 & 60 & 4  & 2*500 & 0.05 \\ 
                 \hline
                \end{tabular}
                \end{center}
            \end {table}

\begin{figure}[h!]
    \centering
    \includegraphics[width=\linewidth]{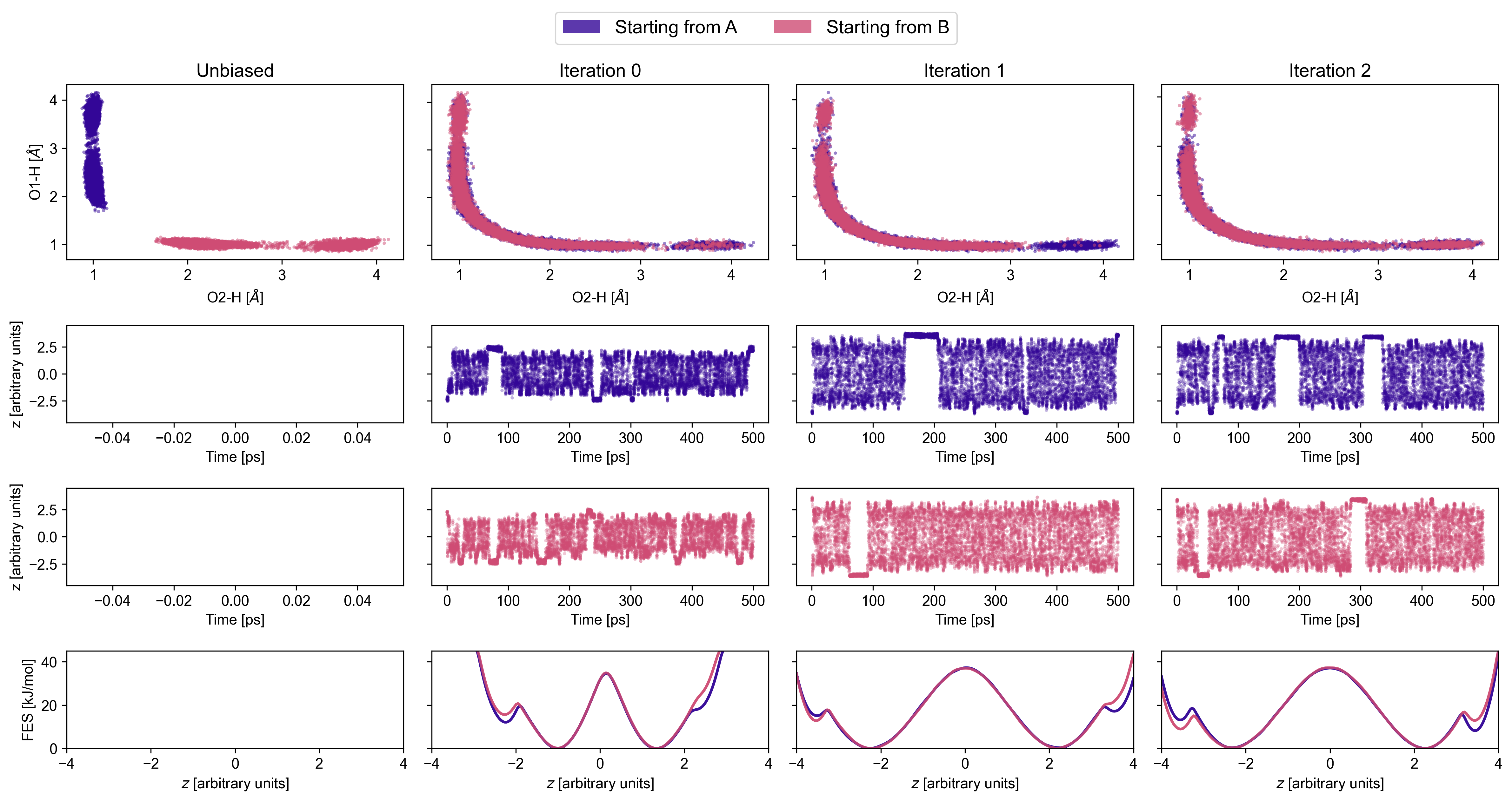}
    \caption{\textbf{Iterative sampling on tropolone.} Top row:  Scatter plot of the sampled points in the space defined by the O2-H and O1-H distances at successive iterations. Points from trajectories initialized in state A are shown in purple, while those initialized in state B are shown in pink.
        Second row: Time series of the $z$ CV for trajectories started from state A.
        Third row: Time series of the $z$ CV for trajectories started from state B.
        Bottom row: Free energy surface (FES) plotted along the $z$ CV, estimated from the sampled configurations at each iteration.}
    \label{SI_fig:tropolone_sampling}
\end{figure}



\clearpage
\section{OAMe-G2 binding}
    \subsection{Computational details}
        \paragraphtitle{Simulations details}
            We performed the simulations with GROMACS-2024.5~\cite{abraham2015gromacs} in combination with the PLUMED-2.9.4~\cite{plumed2019promoting} plugin. We use the GAFF~\cite{wang2004development} force field with RESP~\cite{bayly1993well} charges and the TIP3P~\cite{jorgensen1983comparison} water model. 
            We set the timestep to 2~fs, targeting the NVT ensemble at 300~K via a velocity rescale thermostat~\cite{bussi2007velocity} with a time constant of 0.1~ps.
            The simulation box is cubic with a side of 40.27~$\si{\angstrom}$ and it contains 2100 water molecules in solution together with the host OAMe and the G2 guest molecule. 
            Sodium ions are included to counterbalance excess charges. 
            At every simulation step, the coordinates are aligned so that the vertical axis of the box coincides with the binding axis $h$, and the simulation box is centered on the virtual atom V1.

        \paragraphtitle{The funnel restraint}
        \label{sup_sec:funnel_restraint}
            In our simulations, we used a funnel restraint~\cite{limongelli2013funnel} equivalent to the one previously employed in previous works~\cite{rizzi2021role,bhakat2017resolving,perez2019local} on the same system.
            Here, we summarize the details of such a restraint, while more details can be found in the original works.
            The funnel limits the space available to the ligand in the unbound state U by confining it to a cylindrical volume above the binding site. 
            As the ligand approaches the binding site, the funnel restraint becomes wider so that its presence does not affect the binding process itself. 
            Having aligned with PLUMED the system to a reference configuration where the binding axis is found along the vertical axis, we define $h$ as the projection on the binding axis of the center of the carbon atoms of each ligand and $r$ as its radial component.
            When $h>10 \si{\angstrom}$, the funnel surface is a cylinder with radius $R_{cyl} = 2 \si{\angstrom}$ with its axis along the vertical direction. 
            When $h<10 \si{\angstrom}$, the funnel opens into an umbrella-like shape with a 45 degree angle whose surface is defined by $r=12-h$.
            The force that, for a displacement x, pushes the ligand away from the funnel’s surface is harmonic $-k_Fx$ with $kF=20 $ kJ/mol $\si{\angstrom}^{-2}$. 
            A further harmonic restraint is applied on $h$  to prevent the ligand from getting too far from the host, reaching the upper boundary of the simulation box. 
            The corresponding force is $-k_U(h-18)$ for $h > 18\si{\angstrom}$ and $k_U=40 $ kJ/mol $\si{\angstrom}^{-2}$.

            During training, we set boundaries further to state U so that the labeled configurations used to impose the boundary conditions will not influence the committor training in the subsequent iterative simulations.
            We apply the funnel restraint described above and two additional harmonic restraints $-k_U(h-20)$ for $h > 20 \si{\angstrom}$ and $-k_U(h-18)$ for $h < 18 \si{\angstrom}$, with $k_U = 20$ kJ/mol.
        
            Because of the funnel presence, the free energy difference between the bound and the true unbound state that we extract from enhanced sampling simulations needs a correction that can be calculated as:
                \begin{equation}
                    \Delta G = -\frac{1}{\beta} \log \left( C_0 \pi R_{\text{cyl}}^2 \int_B dh \exp\left[-\beta \left(W(h) - W_U\right)\right] \right)
                \end{equation}
            where $\beta$ = 1/($k_B$T), $C_0$ = 1/1660 $\si{\angstrom}^{-3}$ is the standard concentration, $h$ is the coordinate along the funnel’s axis, $W(h)$ is the free energy along the funnel axis and $W_U$ its reference value in state U. 
            More precisely, we define $W_U$ as the average free energy value in the interval 1.6  $\si{\angstrom}$ < $h$ < 1.8  $\si{\angstrom}$.
            The integral is computed over the state B region that we define as 0.3  $\si{\angstrom}$ < $h$ < 0.8  $\si{\angstrom}$.

            \paragraphtitle{Approximated committor model training details} 
            To model the committor function $q_\theta(\textbf{x})$ at each iteration, we used the same water coordination numbers used in Ref.~\citenum{rizzi2021role}, 8 with respect to 2.5~\AA-spaced virtual atoms (V) along the binding axis and 4 with respect to 4 atoms on the ligand molecule (M) (2 on the ring (1 and 2), 2 on the terminal atoms (3 and 4)) as inputs of a neural network (NN) with architecture [12, 24, 8, 1] nodes/layer.
            For the optimization, we used the ADAM optimizer with an initial learning rate of $10^{-3}$ modulated by an exponential decay with multiplicative factor $\gamma=0.99995$.
            The training was performed for 5000 epochs. 
            The $\alpha$ hyperparameter in the loss function was progressively diminished through the iterations, i.e., 10$^3$, 10$^1$, 10$^{-3}$, and we optimized the $\log$ of the total loss for numerical stability. 
            The number of iterations, the corresponding dataset size, and the $\lambda$ and the OPES \texttt{BARRIER} used in the biased simulations with the corresponding model are summarized in Table~\ref{sup_tab:calixarene_iteration} alongside the simulation time $t_s$ and the output sampling time $t_o$.

            \begin {table}[h!]
                \caption {\textbf{Summary calixarene.} Summary of the iterative procedure for calixarene.} \label{sup_tab:calixarene_iteration}
                \begin{center}
                \begin{tabular}{ |c|c|c|c|c|c|c| } 
                 \hline
                 Iteration & Dataset size & OPES \texttt{BARRIER} [kJ/mol] & $\lambda$ & $t_s$ [ns] & $t_o$ [ps] \\ 
                 \hline
                    0   & 4000  & 50& 3.6  & 2*100 & 1\\
                    1   & 42000 & 50& 3.6  & 2*100 & 1\\
                    2   & 74000 & 50& 3.6  & 2*250 & 1\\
                 \hline
                \end{tabular}
                \end{center}
            \end {table}

    \newpage
    \subsection{Additional figures}
        \begin{figure}[h!]
            \centering
            \includegraphics[width=\linewidth]{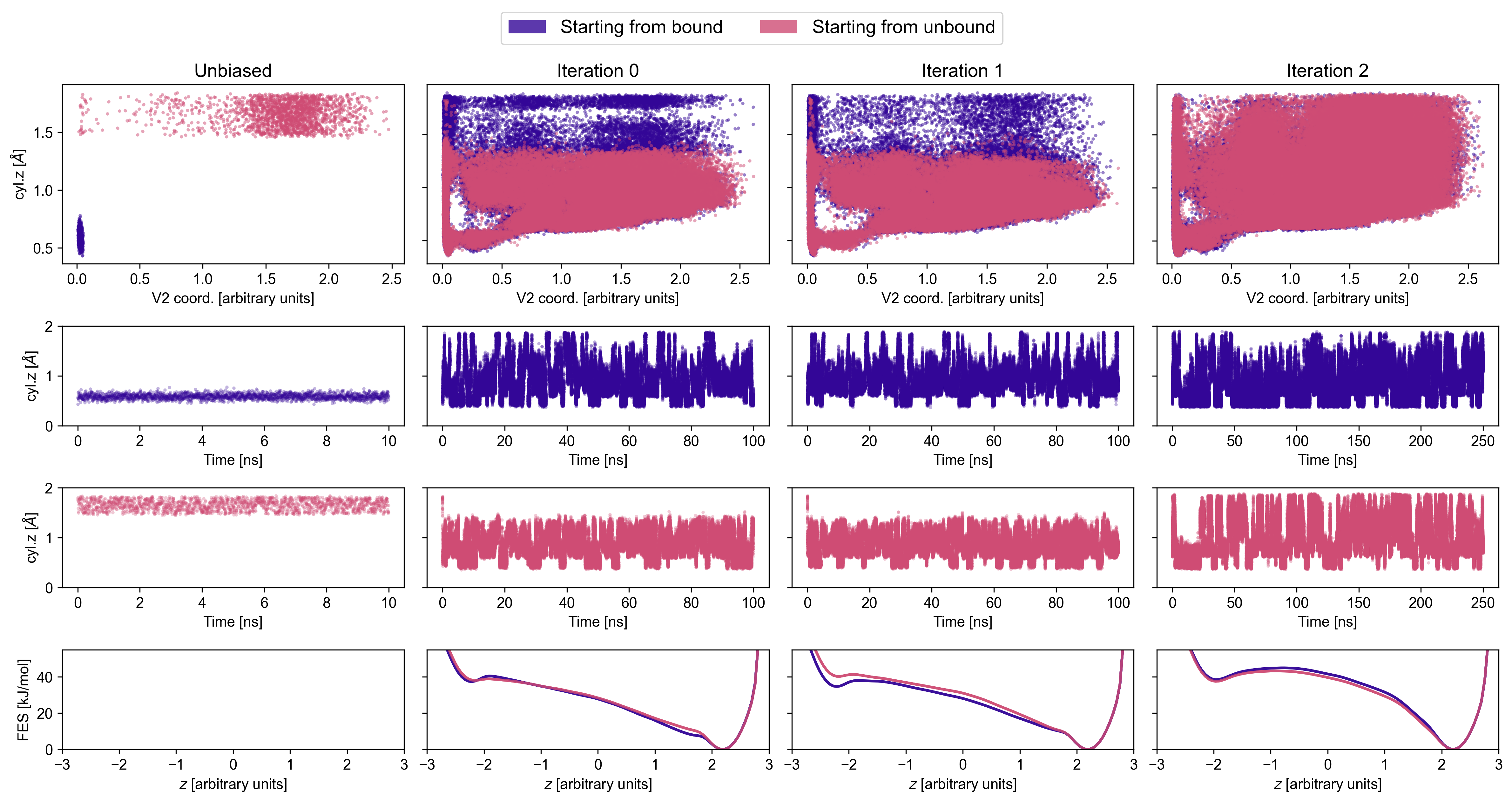}
            \caption{\textbf{Iterative sampling on calixarene.} Top row:  Scatter plot of the sampled points in the space defined by the distance from the binding pocket and the hydration of the binding pocket, at successive iterations. Points from trajectories initialized in the bound state are shown in purple, while those initialized in the unbound state are shown in pink.
        Second row: Time series of the distance from the binding pocket for trajectories started from state A.
        Third row: Time series of the distance from the binding pocket for trajectories started from state B.
        Bottom row: Free energy surface (FES) plotted along the $z$ CV, estimated from the sampled configurations at each iteration.}
            \label{SI_fig:calixarene_sampling}
        \end{figure}



\clearpage
\section{Silicon crystallization}
    \subsection{Computational details}
        \paragraphtitle{Simulations details}
        Silicon crystallization simulations were carried out targeting the NPT ensemble using the same setup of Ref.\cite{bonati2021deep}, employing LAMMPS~\cite{lammps} (22Jul2025 update1) patched with PLUMED-2.9.0.~\cite{plumed2019promoting}
        As an interatomic potential we used the Stillinger–Weber interatomic potential.~\cite{stillinger1985computer}.
        A 3 × 3 × 3 supercell (216 atoms) was simulated in the isothermal–isobaric (NPT) ensemble with a time step of 2~fs. 
        The velocity-rescaling thermostat~\cite{bussi2007velocity} with a target temperature of 1700~K and a relaxation time of 100~fs was used, while the barostat~\cite{martyna1994constant} target pressure was set to 1~atm and the relaxation time to 1~ps.
        First, two 2~ns-long simulations of standard MD in the solid and liquid states were performed. 
        During the simulation, we monitored the fraction of diamond-like atoms computed in PLUMED~\cite{plumed2019promoting} with the Environment Similarity CV,~\cite{piaggi2019phase} with parameters \texttt{SIGMA} = 0.4, \texttt{LATTICE\_CONSTANTS} = 5.43, \texttt{MORE\_THAN} = $\{$ \texttt{R\_0} = 0.5 \texttt{NN} = 12 \texttt{MM} = 24$\}$.
        
        \paragraphtitle{Approximated committor model training details}       
        To model the approximated committor, we used the 95 three-dimensional structure factor peaks as inputs of a neural network (NN) with architecture [95, 48, 24, 1] nodes/layer and an initial normalization layer.
        OPES+$V_K$ simulations biasing the committor-based $z$ variable, with \texttt{PACE} = 500, adaptive sigma with \texttt{ADAPTIVE\_SIGMA\_STRIDE}=500 were performed. 
        For the committor training, we discarded the first part of the simulations in which the bias was stabilizing.
        For the optimization, we used the ADAM optimizer with an initial learning rate of $10^{-3}$ modulated by an exponential decay with multiplicative factor $\gamma=0.99995$.
        The training was performed for 5000 epochs during the first iteration and for 40000 epochs during the following ones. 
        The $\alpha$ hyperparameter in the loss function was progressively diminished through the iterations, i.e., 10$^3$, 10$^-3$, 10$^{-6}$, and we optimized the $\log$ of the total loss for numerical stability. 
        The number of iterations, the corresponding dataset size, and the $\lambda$ and the OPES \texttt{BARRIER} used in the biased simulations are summarized in Table~\ref{sup_tab:silicon_iteration} alongside the simulation time $t_s$ and the output sampling time $t_o$.
        In this case, only data from the previous iteration were used for the training.

        \begin {table}[h!]
                \caption {\textbf{Summary silicon.} Summary of the iterative procedure for silicon crystallization.} \label{sup_tab:silicon_iteration}
                \begin{center}
                \begin{tabular}{ |c|c|c|c|c|c|c| } 
                 \hline
                 Iteration & Dataset size & OPES \texttt{BARRIER} [kJ/mol] & $\lambda$ & $t_s$ [ns] & $t_o$ [ps] \\ 
                 \hline
                    0   & 10000  & 1100& 100  & 2*50 & 1\\
                    1   & 110000 & 1100& 1000  & 2*100 & 1\\
                    2   & 110000 & 1100& 1000  & 2*100 & 1\\
                 \hline
                \end{tabular}
                \end{center}
            \end {table}
        
\begin{figure}[h!]
    \centering
    \includegraphics[width=\linewidth]{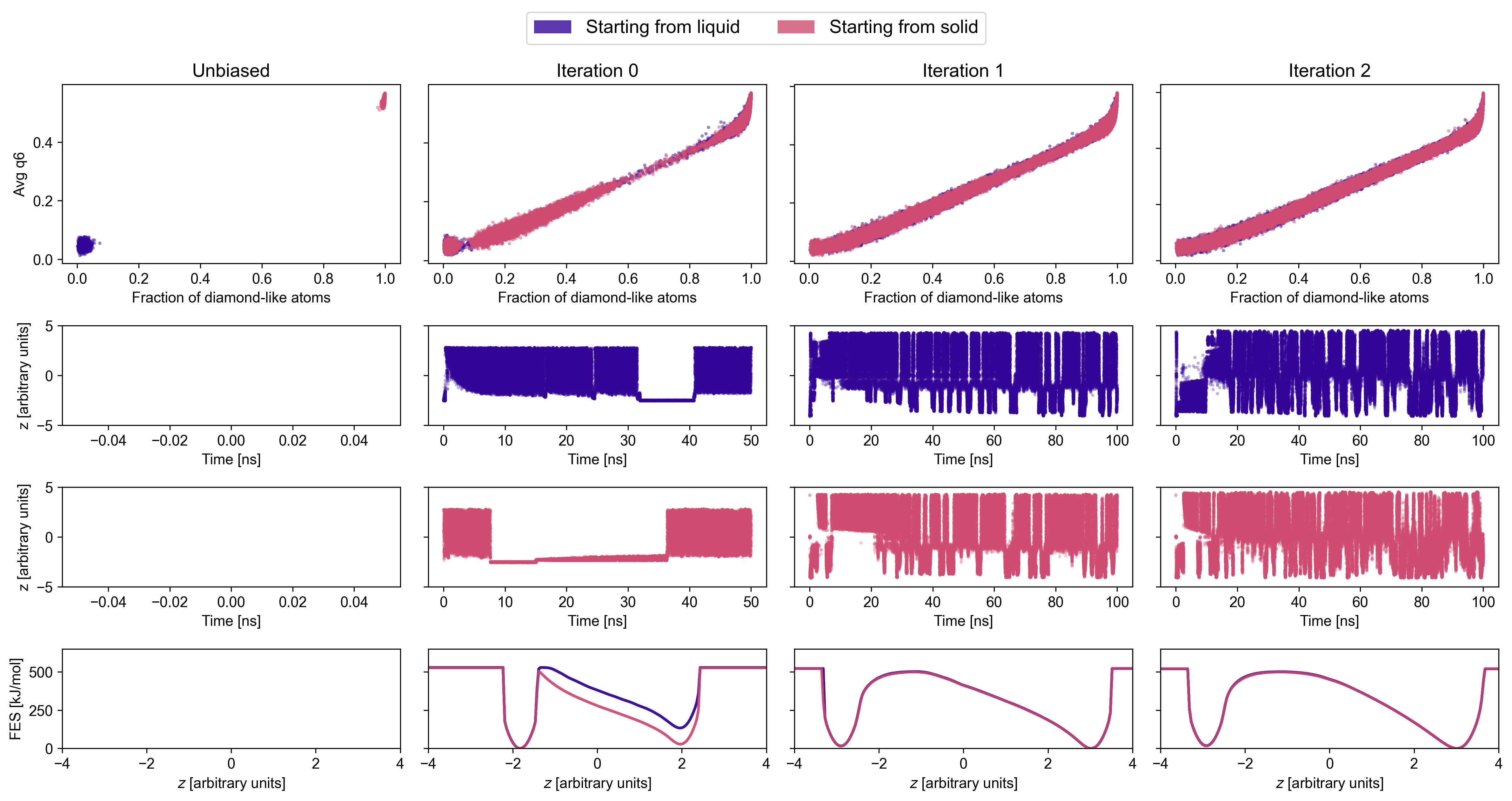}
    \caption{\textbf{Iterative sampling on silicon.} Top row:  Scatter plot of the sampled points in the space defined by the average sixth-order Steinhardt parameter $q6$ and the fraction of diamond-like atoms at successive iterations. Points from trajectories initialized in state A are shown in purple, while those initialized in state B are shown in pink.
        Second row: Time series of the $z$ CV for trajectories started from liquid.
        Third row: Time series of the $z$ CV for trajectories started from solid.
        Bottom row: Free energy surface (FES) plotted along the $z$ CV, estimated from the sampled configurations at each iteration.}
    \label{SI_fig:silicon_sampling}
\end{figure}

\begin{figure}[h!]
    \centering
    \includegraphics[width=0.8\linewidth]{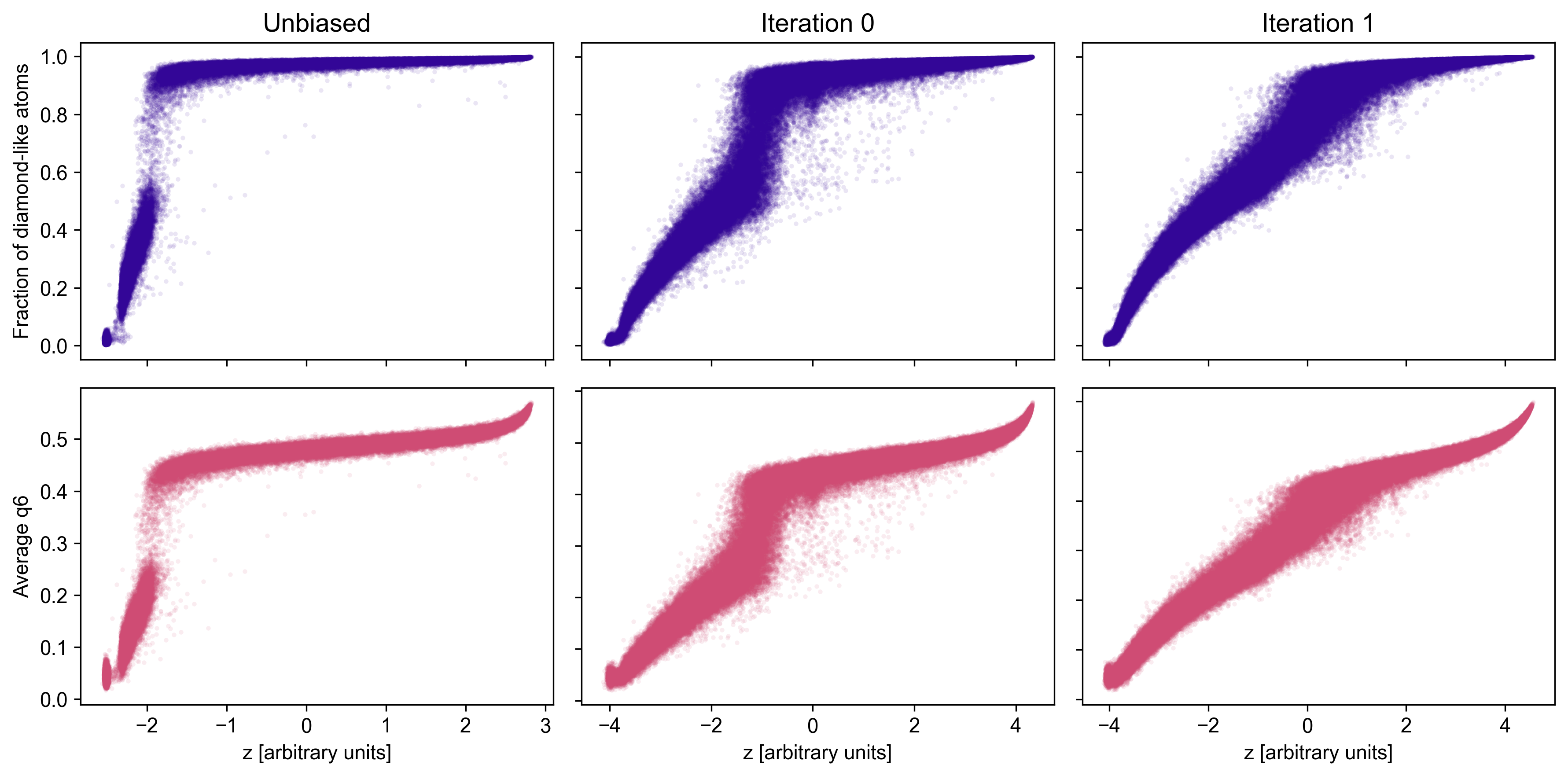}
    \caption{\textbf{Learned models vs reference CVs.} Comparison between the learned $z$ variable at successive iterations and reference CVs, i.e., the fraction of diamond-like atoms (top row) and the average sixth-order Steinhardt parameter (bottom row).}
    \label{SI_fig:silicon_comparison}
\end{figure}



\clearpage
\section*{Supporting References}
\putbib
\newpage

\end{bibunit}

\end{document}